\documentclass[aps,prd,a4paper,preprintnumbers,amsmath,amssymb,superscriptaddress,twocolumn,floatfix,showpacs,nofootinbib]{revtex4}
\usepackage{amssymb}
\usepackage{float,graphicx}

\newcommand{\dd}{{\rm d}}

\newcommand\mpl{M_{\rm Pl}}

\begin{document}
\title{$f(R)$ Gravity and Chameleon Theories}

\author{Philippe~Brax}
\affiliation{ Institut de Physique Th\'eorique CEA, IPhT, F-91191
Gif/Yvette, France, CNRS, URA 2306, F-91191, Gif/Yvette, France.}
\author{Carsten van de Bruck}
\affiliation{ Department of Applied Mathematics, The University of Sheffield,
 Hounsfield Road, Sheffield S3 7RH, United Kingdom}
\author{Anne-Christine Davis}
\affiliation{Department of Applied Mathematics and Theoretical Physics,
Centre for Mathematical Sciences,  Cambridge CB2 0WA, United Kingdom}
\author{Douglas J. Shaw}
\affiliation{Astronomy Units, Queen Mary, University of London, Mile End Road, London E1 4NS, United Kingdom}
\begin{abstract}
We analyse $f(R)$ modifications of Einstein's gravity as  dark
energy models in the light of their connection  with chameleon
theories. Formulated as scalar-tensor theories, the $f(R)$
theories imply the existence of a strong coupling of the scalar
field to matter. This would violate all experimental gravitational
tests on deviations from Newton's law. Fortunately, the existence
of a matter dependent mass and a thin shell effect allows one to
alleviate  these constraints. The thin shell condition also
implies strong restrictions on the cosmological  dynamics of the
$f(R)$ theories. As a consequence, we find that the equation of
state of dark energy  is constrained to be extremely close to $-1$
in the recent past. We also examine the potential effects of
$f(R)$ theories in the context of the E\"ot-wash experiments. We
show that the requirement of a thin shell for the test bodies is
not enough to guarantee a null result on deviations from Newton's
law.
As long as dark energy accounts for a sizeable fraction of the
total energy density of the Universe, the constraints which
we deduce also forbid any measurable deviation of the dark energy
equation of state from -1. All in all, we find that both
cosmological and laboratory tests imply that $f(R)$ models are
almost coincident with a $\Lambda$CDM model at the background
level.
\end{abstract}

\maketitle
\section{Introduction}
The acceleration of the Universe expansion was discovered ten
years ago and is still a deep mystery (see e.g. \cite{de1} for recent results
on observations of dark energy and e.g. \cite{durrer,de2} for theoretical overviews).
Two types of approaches have been considered. One can either introduce a new kind of
matter whose role is to trigger acceleration or modify the
behaviour of gravity at cosmological distances. In the first approach, dark energy is a
new energy form, with all its well-known puzzles such as the cosmological constant problem,
the coincidence problem and the value of the equation of state. This approach is
subject of intense experimental investigation and any deviation
from -1 would be a smoking gun for new physics beyond the standard
models of particle physics and cosmology. On the other hand, in the second approach,
various attempts to modify gravity have been presented (see e.g. \cite{Carroll}-\cite{Amendola};
the literature is vast, see \cite{fara} for a recent overview and further references).
Up to now, they are all plagued with various theoretical problems
such as the existence of ghosts or instabilities. In this paper, we will consider a
modification of Einstein's gravity, the so--called $f(R)$ theories, which do
not seem to introduce any new type of matter and can lead to late
time acceleration. In fact, these theories can be reformulated in
terms of scalar-tensor theories with a fixed coupling of the extra
scalar degree of freedom to matter. As theories of dark energy,
they suffer from the usual problems and are also potentially ruled
out by gravitational tests of Newton's law.

The only way-out for these models is to behave as chameleon
theories \cite{chamKA}, i.e. develop an environment dependent mass
\cite{navarro,brookfield,faulkner,li}. When the density of the
ambient matter in which the scalar field/chameleon propagates is
large enough, its mass becomes large and the smallness of the
generated fifth force range is below the detectability level of
gravitational experiments. On the other hand, for planetary orbits
or any other situations in which gravity is at play in a sparse
environment, one must impose the existence of the so--called thin
shell effect. In this case, the fifth force is attenuated as the
chameleon is trapped inside very massive bodies (the Sun for
instance). It has been argued that the existence of thin shells is
usually enough to salvage $f(R)$ models \cite{navarro,faulkner}.
We show that thin shells do not always guarantee null results in
experimental tests of Newton's law. We exemplify this fact using
the E\"ot-wash setting and obtain strong constraints on the models
which translate into stringent bounds on the present dark energy
equation of state, preventing any detection of a deviation from -1
in the foreseeable future  $\vert 1+w\vert \le 10^{-4}$, where
$w$ is the equation of state of dark energy in the recent
past. This corroborates a similar bound obtained from the
existence of  thin shell for objects embedded in a super-cluster.
It should be noted that this result holds at the background level.
 For higher red-shifts where the effective dark energy
density fraction, $\Omega_{\rm de}$, may become small (or even
vanish), larger deviations can be present as exemplified in the
models in \cite{tsu1,tsu2}  where the equation of state can
deviate from -1 for red-shifts of order $z=2-3$.  In all these
models however $\vert 1 + w\vert \Omega_{\rm de} \ll 1$, and so
even if $w$ deviates significantly from  $-1$, deviations of the
homogeneous cosmology from $\Lambda$CDM are still very small. Detectable deviations from $\Lambda$CDM are envisageable  at the
perturbative level as the growth factor is anomalous at small
scales (see e.g. \cite{green} for a discussion of this point for
the original chameleon model). Some consequences of this fact on
the matter power spectrum and the CMB spectrum of $f(R)$ models
have been presented in Ref. \cite{Hu1,spergel,Hu2}.

The paper is organized as follows: In the subsequent section, we review $f(R)$
models and chameleon theories. In section III we derive the cosmological
thin shell bound on the equation of state. In section IV, we consider tests
of the inverse square law. Finally, we apply these considerations
to particular models in section V. The appendices contain some
technical details.

\section{$f(R)$ Gravities and Chameleon Theories}
\subsection{$f(R)$ theories}
An $f(R)$ theory is a modified gravity theory in which the usual
Einstein-Hilbert Lagrangian for General Relativity, i.e. $R$, is
replaced by some arbitrary function of the scalar curvature i.e.
$f(R)$.  The action for an $f(R)$ gravity theory therefore takes
the following form:
\begin{equation}
S_{f(R)} = \int {\rm d}^4\,x \,\sqrt{-g}\frac{\mpl^2}{2}f(R) + S_{\rm matter}[g_{\mu \nu},\Psi_{i}], \label{action}
\end{equation}
where the $\Psi_{i}$ represent the matter fields.



In this work we are concerned only with metric $f(R)$ theories, in which
only the metric $g_{\mu\nu}$ is the independent variable in the gravitational
sector. The quantity $\Gamma_{\mu \nu}^{\rho}$ is taken to be the
Levi-Civita connection associated with the metric $g_{\mu \nu}$. In these
metric $f(R)$ theories the field equations are:
\begin{eqnarray}
R_{\mu \nu} f^{\prime}(R) &&- \frac{1}{2} f(R) g_{\mu \nu} =  \kappa T^{\rm matter}_{\mu \nu} \label{fReqn1} \\
&& + \nabla_{\mu}\nabla_{\nu} f^{\prime}(R) - g_{\mu \nu} \square f^{\prime}(R). \nonumber
\end{eqnarray}
where $\kappa = 1/\mpl^2$.

\subsection{Transformation to a Scalar-Tensor theory}
 Eq. (\ref{fReqn1}) gives a set of equations which are second order in derivatives of $R$, which is itself second order in derivatives of $g_{\mu \nu}$, making the
 field equations fourth order in $g_{\mu \nu}$.  Finding solutions to fourth order equations can be mathematically and physically troublesome, but fortunately metric $f(R)$ theories can be recast as a scalar tensor theory with only second order equations via a well known conformal transformation.
We define $\phi$ by
\begin{equation}
\exp\left(-\frac{2\beta \phi}{\mpl}\right) = f^{\prime}(R), \nonumber
\end{equation}
where $\beta = \sqrt{1/6}$. We also define the \emph{Einstein
frame} metric $\bar{g}_{\mu\nu}$ by a conformal transformation
\begin{equation}
\bar{g}_{\mu \nu} = e^{-\frac{2\beta \phi}{\mpl}} g_{\mu \nu}, \nonumber
\end{equation}
and let $\bar{R}$ be the scalar curvature of $\bar{g}_{\mu \nu}$.  When rewritten in terms of $\bar{g}_{\mu \nu}$ and $\phi$, Eq. (\ref{action}) becomes:
\begin{eqnarray}
S_{\rm ST} = &&\int {\rm d}^4\,x\, \sqrt{-\bar{g}}\left(\frac{\mpl^2}{2}\bar{R} - \frac{1}{2}\bar{g}^{\mu \nu}\nabla_{\mu} \phi \nabla_{\nu} \phi - V(\phi)\right)\nonumber \\ &&+ S_{\rm matter}[e^{\frac{2\beta \phi}{\mpl}}\bar{g}_{\mu \nu},\Psi_{i}], \label{action2}
\end{eqnarray}
where the potential $V(\phi)$ is given by:
\begin{equation}
V(\phi) = \frac{\mpl^2\left(R f^{\prime}(R) -f(R)\right)}{2 f^{\prime}(R)^2}. \label{Veqn}.
\end{equation}
When the action is written in the form of Eq. (\ref{action2}), we say that we are working in the Einstein frame.  The field equations then become:
\begin{eqnarray}
\bar{G}_{\mu \nu} &=& \bar{R}_{\mu \nu} - \frac{1}{2}\bar{R}\bar{g}_{\mu \nu} \label{feqn2} = \kappa \bar{\nabla}_{\mu}\phi \bar{\nabla}_{\nu}\phi \\ &&- \kappa\bar{g}_{\mu \nu} \left[\frac{1}{2}(\bar{\nabla} \phi)^2 + V(\phi)\right] +\kappa T^{\rm matter}_{\mu \nu}  \nonumber \\
\bar{\square} \phi  &=&  V^{\prime}(\phi) - \frac{\beta}{\mpl} T^{\rm matter}. \label{eqnPhi}
\end{eqnarray}
In the above and subsequent expressions, the covariant
derivatives, $\bar{\nabla}_{\mu}$, obey $\bar{\nabla}_{\mu}
\bar{g}_{\mu \nu=0}$ and all indices are raised and lowered with
$\bar{g}_{\mu \nu}$ unless stated otherwise.  We note that in the
Einstein frame $T^{\rm matter}_{\mu \nu}$ is not conserved but
instead:
\begin{equation}
\bar{\nabla}_{\mu} T^{{\rm matter}\,\mu}{}_{\nu} =  \frac{\beta}{\mpl} T^{\rm matter} \bar{\nabla}_{\nu} \phi. \label{eqnT}
\end{equation}
This implies that matter will generally feel a new or `fifth'
force due to gradients in $\phi$.
   We note from Eq. (\ref{feqn2}) that, when written as a scalar tensor theory, gravity in an $f(R)$ theory is essentially General Relativity, and all the modifications are essentially due to the effective `fifth force' and to the energy density of $\phi$.  Much of our intuition for how gravity works is based on how it works in General Relativity. When an $f(R)$ theory is written as a scalar tensor theory we can readily make use of this intuition in solving the field equations.  This may not be the case, however, in the original frame in which the equations were fourth order and so in those circumstances one would have to be more careful. Note that all physical observables must be independent of the choice of frame, i.e. the choice of metric $g_{\mu \nu}$ or $\bar{g}_{\mu \nu}$.
\subsection{Chameleon Theories}
Since $f(R)$ theories are equivalent to scalar tensor theories one
can  generally directly apply the plethora of constraints on
scalar tensor models.  In particular, since the extra degree of
freedom, $\phi$, couples to matter with gravitational strength,
tests of the inverse square law such as the E\"{o}t-Wash
experiment require that $\phi$ have a mass, $m_{\phi}=
\sqrt{V_{,\phi\phi}}$, greater than  $1\,{\rm meV}$.
Cosmologically $\phi$ would then have been fixed at its minimum
since very early times, and physics over astrophysical scales
would be indistinguishable from predicted by unmodified General
Relativity with a cosmological constant.  Both the coincidence
problem and the problem of the small size of the cosmological
constant would not be alleviated in this scenario. However, this
is not the whole story.  Laboratory constraints on scalar tensor
theories can be greatly relaxed if $m_{\phi}=
\sqrt{V_{,\phi\phi}}$ develops a strong dependence on the ambient
density of matter.  Theories in which such a dependence is
realized are said to have a chameleon mechanism and to be
chameleon theories. In such theories, $\phi$ can be heavy enough
in the environment of the laboratory tests so as to evade them,
whilst remaining relatively light on cosmological scales.  It must
be stressed that even with a chameleon mechanism, it is still very
difficult, if not impossible, to construct such a theory where the
late time cosmology would be observational distinguishable from
the usual $\Lambda$CDM model.  To the best of our knowledge all
such theories which are also experimentally viable require a
fairly high degree of fine tuning to ensure that the effective
cosmological constant is small enough.

A chameleon theory is essentially just a scalar-tensor theory in
which the potential has certain properties.  As such Eqs.
(\ref{action2} - \ref{eqnT}) also define a chameleon theory for
certain classes of $V(\phi)$.  In these circumstances the $f(R)$
theory would be equivalent to a chameleon theory.  In a general
chameleon theory, $\beta$, which parametrizes the strength of the
coupling of $\phi$ to matter, could take any value and potentially
even be different for different matter species.  If a chameleon
theory is equivalent to a $f(R)$ theory, however, $\beta$ is fixed
to be $\sqrt{1/6}$ and is the same for all types of matter. If a
$f(R)$ theory is not equivalent to a chameleon theory  it would be
generally ruled out by laboratory tests of gravity and / or result
in no detectable deviations from General Relativity over
astrophysical scales.

For an $f(R)$ theory to have a chameleon mechanism one must
require that, in at least some region of $\phi$-space:
\begin{equation}
V^{\prime}(\phi) < 0,\qquad V^{\prime \prime}(\phi)> 0,\qquad  V^{\prime \prime \prime}(\phi) < 0. \nonumber
\end{equation}
It follows from the definition of $\phi$ that:
\begin{equation}
\frac{\dd \phi}{\dd R} = -\frac{\mpl}{2\beta} \frac{f^{\prime
\prime}}{f^{\prime}}.
\end{equation}
and therefore the derivatives follow
\begin{eqnarray}
V^{\prime}(\phi) &=& \frac{\beta \mpl}{f^{\prime\, 2}}\left[R f^{\prime}- 2f\right], \\
V^{\prime \prime}(\phi)&=& \frac{1}{3}\left[\frac{R}{f^{\prime}}+\frac{1}{f^{\prime \prime}} - \frac{4f}{f^{\prime\,2}}\right],\\
V^{\prime \prime \prime}(\phi) &=& \frac{2\beta}{3\mpl}
\left[\frac{3}{f^{\prime \prime}} + \frac{f^{\prime}f^{\prime
\prime \prime}}{f^{\prime \prime\,3}}+\frac{R}{f^{\prime}} -
\frac{8f}{f^{\prime\,2}}\right]
\end{eqnarray}
In general, this gives strong constraints on the form of $f(R)$.
In the following we will study examples where these conditions are
met.

When these conditions are satisfied, the mass of $\phi$ in a
suitable large region with density $\rho$ will increase with
$\rho$.  In order to evade constraints coming from local tests of
gravity, it is not, however, enough that a theory possess a
chameleon mechanism; the mechanism must, in addition be strong
enough for chameleonic behaviour to occur for the test masses used
in the laboratory gravity experiments.

\subsection{Thin-Shells}
\subsubsection{Chameleon Theories}
Chameleon theories do not behave like linear theories of massive
scalar fields. In situations where massive bodies are involved,
the chameleon field is trapped inside such bodies and its influence
on other bodies is only due to a thin shell at the outer edge of a
massive body\cite{chamKA}. As a result, the field outside the massive body for distances less than
the range of the chameleon force in the outer vacuum is
effectively damped leading to a shielded fifth force which becomes
undetectable. The criterion for a thin shell is
\begin{equation}
\frac{\Delta \phi}{m_{\rm Pl}} \le 2\beta\Phi_N
\end{equation}
where $\Delta\phi=\phi_\infty -\phi_0$ is the field difference
from far inside the body to very far away. We define the body and the region outside it to have densities $\rho_{0}$ and $\rho_{\infty}$ respectively. It involves Newton's
potential $\Phi_N$ at the surface of the body. In general the
field values at infinity, $\phi_{\infty}$, and deep inside, $\phi_{0}$, are related to $\rho_{\infty}$ and $\rho_{0}$ by
\begin{equation}
\partial_\phi V= -\beta \frac{\rho}{m_{\rm Pl}}.\label{min}
\end{equation}
In most current situations involving runaway potentials, when $\rho_{0} \gg \rho_{\infty}$, this implies that $\phi_\infty \ge \phi_0$. Hence, $\Delta \phi=
\phi_\infty$ implying that cosmological information can be
inferred from local tests. Moreover, in a cosmological setting,
the chameleon sits at the minimum (\ref{min}) during the matter
era. As a result, the variation of the equation of state in the
recent past is severely constrained. Another important consequence
of the chameleon effect is the existence of an anomalous growth of
the density contrast for scales lower than the inverse mass of the
chameleon, i.e. it grows like $a^\nu$ where $\nu\approx
\frac{-1+\sqrt{1+24(1+2\beta^2)}}{2}$\cite{green}. In the $f(R)$ setting, some of the consequences of this anomalous growth on the CMB and the matter power spectrum
have been analysed
using the convenient variable
\begin{equation}
B=\frac{f_{RR}}{f_R} \frac{dR}{dH}H
\end{equation}
whose square root  represents the compton wave-length, i.e. the
inverse mass of the chameleon, in horizon units
$H^{-1}$\cite{Hu1,spergel,Hu2}. Effects on structure formation
could be seen for values as low as $B=10^{-4}$ in future galaxy
surveys\cite{spergel}. In the following we will find an explicit
example of logarithmic $f(R)$ model which could lead to effects on
scales as large as 100 $h^{-1}{\rm Mpc}$. All these  facts
will be crucial in the following.

\subsubsection{Thin shells in the language of $f(R)$ theories}

It is useful to write the function $f(R)$ in the form $f(R)= R +h(R)$,
where $h$ measures the deviation to Einstein gravity. To leading order, as a consequence of Ref. \cite{chamKA}, the
thin shell condition can be formulated as \cite{Hu2} 
\begin{equation}
|\Delta h'(R)| \le \frac{2}{3} \Phi_N.
\end{equation}
As Newton's potential is small on cosmological scales, with an
upper bound around $10^{-4}$, this implies that $h'$ must have
very small variations. The thin shell condition is a constraint on
local experiments at the present time. It has nothing to say, a priori, about the
evolution of the universe since matter equality for instance.
Another useful combination (which is not to be confused with the chameleon mass $m_{\phi}$) has been used
\begin{equation}
m= \frac{Rh''(R)}{1+h'(R)}
\end{equation}
It has been shown that the existence of a matter era followed by
an accelerated period requires $m<0.1$. For models where $m$ is
(nearly) a power law, the thin shell constrain implies that $m$ is
much smaller for reasonable powers. In the following, we will
obtain a bound on the equation of state at present time which
implies that departures from $\Lambda$CDM are tiny.

\section{Thin-shell constraints on Cosmology}
\label{sec:cosmo} In subsequent sections we will assume that test
bodies used in laboratory based gravity experiments have
thin-shells. In the absence of any thin-shell, the inverse square
law tests, such as the E\"{o}t-Wash experiment \cite{EotWash} (as
well as other tests of gravity over longer ranges), rule out
theories with $\beta = 1/\sqrt{6}$ as it is in $f(R)$ theories.
The thin-shell requirement must therefore be
satisfied by any physically viable $f(R)$ theory. Although it is
not often appreciated, the thin-shell condition for laboratory
test masses actually places extremely tight constraints on the
recent cosmological evolution of $\phi$.  In this section we
consider those constraints in the context of a general $f(R)$
theory.

In any single field scalar tensor theory there is a choice of
frame. In the Jordan frame, the laws of physics in a local
inertial frame are the same everywhere, however Newton's constant,
$G_{N}$, is different at different points in space and time.  In
the Einstein frame, $G_{N}$ is chosen to be fixed but, as a result,
local particle physics is position dependent. The process of
converting astronomical observations to cosmological parameters
generally involves making assumptions about how today's laws of
particle physics are related to those in the past. This said, if
the relative changes in $G_{N}$ (in the Jordan frame) are small
i.e. $\ll 1$, the differences between cosmological parameters in
the two frames are only very slight.  For instance, to calculate a
redshift, one must compare the observed  wavelength, $\lambda_{\rm
obs}$ of a particular absorption or emission band to the
wavelength that band would have had at emission, $\lambda_{\rm
e}$.  Since one cannot go to the astronomical object in question
and directly observe the wavelength at emission, it is generally
assumed that particles physics in the past obeyed the same laws as
it does today and so replace $\lambda_{\rm e}$ with the wavelength
of the band as it is measured in a laboratory today, $\lambda_{\rm
today}$.   When one is dealing with scalar-tensor theories, the
assumption that $\lambda_{\rm e}=\lambda_{\rm today}$ is
equivalent to a choice of frame, specifically the Jordan frame.

To make comparison with observations straight-forward, one should
therefore quote cosmological parameters for the Jordan frame. This
said, it is often more straightforward to perform calculations in
the Einstein frame and then merely express the results in
terms of Jordan frame quantities.

Cosmologically, in the Jordan frame we have:
\begin{equation}
\dd s^2 = a^2(\eta) \left[-\dd \eta^2 + \gamma_{ij}\dd x^{i}\dd x^{j}\right],
\end{equation}
and $\phi$ obeys:
\begin{equation}
-\frac{1}{a^2} \Phi_{,\eta \eta} - 2\frac{a_{,\eta}}{a^3} \Phi_{,\eta} = \frac{\kappa}{3}\left[ T_{\rm matter} + 2\Phi^3 V_{,\Phi} \right] \label{PhiJordan}
\end{equation}
where $\Phi = e^{-2\beta \phi/\mpl} = f^{\prime}(R)$.  At late
times, when it is appropriate to ignore the contribution of
radiation to the total energy density of the Universe, we  have
\begin{equation}
\frac{3a_{,\eta}^2}{a^4} = \frac{\kappa \rho_{\rm matter}}{\Phi} + \kappa \Phi V(\phi) - \frac{3a_{,\eta}\Phi_{,\eta}}{a^3\Phi} - \frac{3k}{a^2}. \label{H1eqn}
\end{equation}
The Einstein equations also give:
\begin{equation}
\frac{2a_{,\eta \eta}}{a^{3}} - \frac{a_{,\eta}^2}{a^4} = \kappa \Phi V(\phi) - \frac{\Phi_{,\eta \eta}}{\Phi a^2} +\frac{2k}{a^2} - \frac{a_{,\eta}\Phi_{,\eta}}{a^3 \Phi}. \label{addot}
\end{equation}
We assume that measurements are interpreted in terms of General
Relativity, where the energy density of the Universe is assumed to
be due to non-interacting, dark energy and normal matter. Thus we
write
\begin{eqnarray}
H^2 = \frac{a_{,\eta}^2}{a^4} &=& \frac{\kappa}{3 \Phi_0} \rho_{\rm matter} + \frac{\kappa}{3 \Phi_0} \rho_{\rm de}^{\rm eff} - \frac{k}{a^2} \label{Heqn2} \\ &=&  \left(\Omega_{\rm m}^{\rm eff} + \Omega_{\rm de}^{\rm eff}\right)H^2-\frac{k^2}{a^2}.\nonumber
\end{eqnarray}
The above equation partly defines $\rho_{\rm de}^{\rm eff}$,
$\Omega_{\rm m}^{\rm eff}$
and $\Omega_{\rm de}^{\rm eff}$; today
$\Phi = \Phi_0$.  Now in the Jordan frame: $\rho_{\rm matter}
\propto a^{-3}$.   If the effective dark energy equation of state
parameter, $w_{\rm eff}^{\rm de}$, were constant it would obey:
$\rho_{\rm de}^{\rm eff} \propto a^{-3(1+w_{\rm de}^{\rm eff})}$.
More generally however the effective dark energy equation of state
is then given by:
\begin{equation}
\rho_{{\rm eff},\eta}^{\rm de} =  -3\frac{a_{,\eta}}{a}(1+w_{\rm eff}^{\rm de})\rho_{\rm eff}^{\rm de}. \label{wdef}
\end{equation}
Taking the $\eta$-derivative of Eq. (\ref{Heqn2}) we get:
\begin{eqnarray}
\left(2\frac{a_{,\eta \eta}}{a^{3}} - 4\frac{a_{,\eta}^2}{a^4}\right)\frac{a_{,\eta}}{a} &&= \frac{\kappa}{3\Phi_0} \rho_{{\rm matter},\eta} \nonumber \\&&+\frac{\kappa}{3\Phi_0}\rho_{{\rm de},\eta}^{\rm eff} + 2\frac{k}{a^2}\frac{a_{,\eta}}{a},\nonumber
\end{eqnarray}
and so using the Eq. (\ref{wdef}) and $\rho_{\rm matter} \propto a^{-3}$ we have:
$$
\frac{2a_{,\eta \eta}}{a^{3}} - \frac{4a_{,\eta}^2}{a^4} - \frac{2k}{a^2} = -\frac{\kappa}{\Phi_0} \rho_{\rm matter} - \frac{\kappa}{\Phi_0}(1+w_{\rm de}^{\rm eff})\rho_{\rm de}^{\rm eff}.
$$
Finally by adding $3H^2$ to both sides and using Eq. (\ref{Heqn2}) we have:
$$
\frac{2a_{,\eta \eta}}{a^{3}} - \frac{a_{,\eta}^2}{a^4} + \frac{k}{a^2} = - \frac{\kappa}{\Phi_0}w_{\rm de}^{\rm eff}\rho_{\rm de}^{\rm eff}.
$$
So by rearranging the Friedmann equations we have found that
\begin{eqnarray}
w_{\rm de}^{\rm eff}\kappa \rho_{\rm de}^{\rm eff}/\Phi_0 &=& -\frac{2a_{,\eta \eta}}{a^3}+\frac{a_{,\eta}^2}{a^4} - \frac{k}{a^2}, \\
&=& \left[ \frac{\Phi_{,\eta \eta}}{\Phi a^2} + \frac{\Phi_{,\eta}a_{,\eta}}{\Phi a^{3}} - \kappa\Phi V(\phi)\right]. \nonumber
\end{eqnarray}
By comparing Eqs. (\ref{H1eqn}) and (\ref{Heqn2}) we see that:
$$
\frac{\kappa}{\Phi_0}\rho_{\rm de}^{\rm eff} = \frac{\kappa}{\Phi_0} \rho_{\rm matter} \left(\frac{\Phi_0}{\Phi} -1\right) + \kappa \Phi V(\Phi)  - 3\frac{a_{,\eta}}{a^3}\frac{\Phi_{,\eta}}{\Phi}.
$$
Therefore,
\begin{eqnarray}
(1+w_{\rm de}^{\rm eff})\kappa \rho_{\rm de}^{\rm eff}/\Phi_0 = &&\frac{\Phi_{,\eta \eta}}{\Phi a^2} - \frac{2\Phi_{,\eta}a_{,\eta}}{\Phi a^3}\\ &&+ \frac{\kappa}{\Phi_0} \rho_{\rm matter} \left(\frac{\Phi_0}{\Phi} -1\right).\nonumber
\end{eqnarray}
Thus, using $3\Omega_{\rm de}^{\rm eff} H^2 = \kappa\rho_{\rm de}^{\rm eff}/\Phi_0$, we have:
\begin{eqnarray}
\left(1+w_{\rm de}^{\rm eff}\right) \Omega_{\rm de}^{\rm eff} &=& \left[\frac{\Phi_{,\eta \eta}}{3\Phi a^2 H^2} - \frac{2\Phi_{,\eta}}{3\Phi a H} \right. \label{weffeqn1} \\ &&\left.  + \left(\frac{\Phi_0}{\Phi}-1\right) \Omega_{\rm m}^{\rm eff}\right].\nonumber
\end{eqnarray}
$\left(1+w_{\rm de}^{\rm eff}\right) \Omega_{\rm de}^{\rm eff} $
parametrizes the magnitude of deviations from $\Lambda$CDM. If
$\phi = -(\mpl/2\beta) \ln \Phi$ has changed by $\Delta \phi$ in
the last Hubble time, Eq. (\ref{weffeqn1}) implies that, in the
recent past and in the Jordan frame, to within an order of
magnitude:
\begin{equation}
\left \vert \left(1+w_{\rm de}^{\rm eff}\right) \Omega_{\rm de}^{\rm eff}\right \vert \sim \mathcal{O}\left( \frac{\beta \vert \Delta \phi\vert}{\mpl}\right). \label{wcon}
\end{equation}
For later use we rewrite Eq. (\ref{weffeqn1}) in terms of $p = \ln
a$:
\begin{eqnarray}
\left(1+w_{\rm de}^{\rm eff}\right) \Omega_{\rm de}^{\rm eff} &=& \frac{\frac{2\Phi_{pp}}{\Phi} + \left(2\Omega_{\rm de}^{\rm eff} - \Omega_{\rm m}^{\rm eff}-4\right)\frac{\Phi_{p}}{\Phi}}{3(2+\frac{\Phi_{p}}{\Phi})}  \label{weffeqn}\nonumber \\
&+& \frac{2\left(\frac{\Phi_0}{\Phi}-1\right) \Omega_{\rm m}^{\rm eff}}{2+\frac{\Phi_p}{\Phi}}.
\end{eqnarray}

In both the Einstein and Jordan frames,
$e^{\frac{\beta(\phi_1-\phi_0)}{\mpl}} - 1$ gives the relative
change in the ratio of any particle mass, $m_{\rm p}$, and the
Planck mass, $\mpl$ between the times when $\phi=\phi_1$ and when
$\phi = \phi_0$.  In the Einstein frame $\mpl$ is constant but
$m_{\rm p}$ varies whereas in the Jordan frame the converse holds;
the ratio of the two masses, being a dimensionless quantity, is
the same in either frame.  WMAP constrains any such variation in
$m_{\rm p}/\mpl$ between now and the epoch of recombination to be
$\lesssim 5\%$ at $2\sigma$ ($\lesssim 23\%$ at $4\sigma$)
\cite{WMAPvaryG}.  It follows that since recombination
\begin{equation}
\vert e^{\frac{\beta  \Delta \phi }{\mpl}}-1\vert < 0.05. \label{WMAPbound}
\end{equation}
Light element abundances provide similar constraints on any
variation in Newton's constant $G_N$ between the present day and
the time of nucleosynthesis \cite{BBNvaryG}.

Thin-shell constraints, however, provide an even tighter bound on the allowed
change in $\phi$.
To consider these constraints we work in the Einstein frame,
however $\Delta \phi$ is the same in either frame.

We assume, as is the case for the real Universe, that the scales
of the inhomogeneous regions are small compared to the horizon
scale, and that the Universe is approximately homogeneous when
coarse-grained over scales larger than some $L_{\rm hom} \ll
H^{-1}$.  Thus over scales larger than $L_{\rm hom}$, $\phi
\approx \phi_{b}(t)$ and since $L_{\rm hom} \ll H^{-1}$, we can
work entirely over sub-horizon scales, which simplifies the
analysis greatly.   We also assume that the curvature of spacetime
is weak over scales smaller than $L_{\rm hom}$. This is equivalent
to assuming that the Newtonian potential, $U$, is small as
are the peculiar velocities, $v^{i}$, of any matter particles, i.e they
are non-relativistic.

Exploiting both the assumption that $H L_{\rm hom} \ll 1$ and that
gravity is weak inside the the inhomogeneous regions i.e. $U\ll 1$
and $v^{i}v^{i} \ll 1$, we write $\phi = \phi_b(t) + \delta \phi$
and have to leading order in the small quantities and over
sub-horizon scales:
$$
\nabla^2 \delta \phi = V^{\prime}(\phi) + \frac{\beta \rho}{\mpl} + \ddot{\phi}_b + 3H \dot{\phi}_b.
$$
Now
$$
-\ddot{\phi}_b - 3H \dot{\phi}_b = V_{,\phi}(\phi_b) + \frac{\beta \rho_b}{\mpl},
$$
and so
$$
\nabla^2 \delta \phi = m_b^2 \delta \phi + \frac{\beta \delta \rho}{\mpl} + A(\phi,\phi_b).
$$
where
$$
A(\phi,\phi_b) \equiv \left[V^{\prime}(\phi_b + \delta \phi) - V^{\prime}_b - m_{b}^2 \delta \phi\right].
$$
Thus
\begin{eqnarray}
\delta \phi = &&-\frac{1}{4\pi} \int \dd^3 x^{\prime} \frac{e^{-m\vert x - x^{\prime}\vert}}{\vert x - x^{\prime}\vert} \left[\frac{\beta \delta \rho(x^{\prime},t)}{\mpl} \right. \nonumber \\ && \left. + A(\phi(x^{\prime},t),\phi_b(t))\right].
\end{eqnarray}
It is straightforward to show that the condition $V^{\prime \prime \prime} < 0$, which must hold for any chameleon theory, implies that $A(\phi,\phi_b) < 0$ for all $\phi$ and $\phi_b$.  Thus
\begin{equation}
\delta \phi > -\delta \phi_1 = -\frac{1}{4\pi} \int \dd^3 x^{\prime} \frac{e^{-m\vert x - x^{\prime}\vert}}{\vert x - x^{\prime}\vert} \frac{\beta \delta \rho(x^{\prime},t)}{\mpl}. \nonumber
\end{equation}
Now if we require that  a test mass at $r=0$ with central density
$\rho_c
> \rho_b$ has a thin-shell, we must impose that at $r=0$,
$\phi \approx \phi_c$, where
$$
V_{,\phi}(\phi_c) = -\frac{\beta \rho_c}{\mpl}.
$$
Thus $\phi$ must be able to change by at least $\phi_c -\phi_b =
-\Delta \phi_{bc} < 0$ i.e. we have the following necessary
condition for thin-shell
\begin{equation}
\frac{\beta\Delta \phi_{bc}}{\mpl} < \frac{\beta\delta \phi_{1}}{\mpl} =  \frac{1}{3} \int \dd^3 x^{\prime} \frac{e^{-m\vert x - x^{\prime}\vert}}{\vert x - x^{\prime}\vert} G\delta \rho(x^{\prime},t).
\end{equation}
The right hand side of this equation is $\mathcal{O}(U/3)$ or
smaller, and the largest values of the  peculiar Newtonian
potential for realistic models of our Universe are roughly $<
10^{-4}$, and are generally around $10^{-6}-10^{-5}$ for large
clusters and superclusters \cite{Hu2}.  Thus we have the following
conservative constraint on the cosmological value of the field
today:
\begin{equation}
\frac{\beta \Delta \phi_{bc}}{\mpl} < 10^{-4}, \label{phicoscon}
\end{equation}

We have defined $f(R) = R + h(R)$.  The thin-shell constraint
certainly  ensures that cosmologically today $\vert \beta \phi
/\mpl \vert \ll 1$ and since we have $1+h^{\prime}(R) =
\exp(-2\beta \phi/\mpl)$ by definition we are therefore justified
in assuming that we have $\vert h^{\prime}(R) \vert \ll 1$.  Then
assuming that  $\vert h^{\prime} \vert \ll 1$ we find that the
potential, $V(\phi)$, is given by:
$$
\kappa V(\phi) \approx \frac{1}{2}\left(Rh^{\prime}(R) -
h(R)\right) ,
$$
and
$$
-\frac{1}{\beta \mpl} V_{,\phi} \approx R(1-2h^{\prime}(R)) +
2h(R) - Rh^{\prime}(R).
$$
To leading order then in $\vert h^{\prime}(R)\vert$ we have:
$$
-\frac{1}{\beta \mpl} V_{,\phi} \approx R - 4\kappa V(\phi).
$$
The chameleon mass squared, $m_{\phi}^2 = V_{,\phi \phi}$ is then
given, to leading order, by:
\begin{equation}
m_{\phi}^2 = V_{,\phi \phi} \approx \frac{1}{3h^{\prime
\prime}(R)}.
\end{equation}
Provided $m_{\phi}^2 / H^2 \ll 1$, then the chameleon field will
remain close to the minimum of its effective potential
\cite{chamcos} cosmological, i.e. $V_{,\phi} = -\beta \rho_{\rm
matter}/\mpl$ and the energy density of the chameleon field will
be dominated by its potential.  Assuming that this is the case we
would have:
$$
R \approx 4\kappa V(\phi) + \kappa \rho_{\rm matter},
$$
and defining $\Omega_{m} = \kappa \rho_{\rm matter}/3H^2$ and
$\Omega_{\rm de} \approx \kappa V(\phi)/3H^2$, we have:
$$
R \approx 3(4\Omega_{\rm de} + \Omega_{\rm m})H^2,
$$
and so $m_{\phi}^2 / H^2 \gg 1$ becomes:
$$
\frac{4\Omega_{\rm de} + \Omega_{\rm m}}{R h^{\prime \prime}(R)}
\gg 1.
$$
Therefore, in many theories, an observationally viable evolution
of $\phi$ requires that it has sat close to the effective minimum
of its potential since recombination \cite{chamcos} i.e.:
$$
V_{,\phi}(\phi_b(t)) \approx -\frac{\beta \rho_{\rm matter}(t)}{\mpl}
$$
Since the background density of matter decreases with time,
$V_{,\phi \phi}>0$ implies that $\phi$ increases with time. Thus
for test mass with density $\rho_c \sim \mathcal{O}(1)\,{\rm
g\,cm}^{-3}$, we have in the recent past, i.e. out to $z \approx
1$:
$$
\phi_c < \phi_{b}(t) < \phi_{b}(t_0)
$$
where $t=t_0$ is the current time.  In this case Eq. (\ref{phicoscon}) gives the following conservative constraint:
$$
\frac{\beta}{\mpl}\left(\phi_b(t_0) - \phi_b(t)\right) < 10^{-4},
$$
and so, from Eq. (\ref{wcon}) we obtain that:
\begin{equation}
\left \vert 1+w_{\rm eff} \right \vert \Omega_{\rm de}^{\rm eff} < 10^{-4}.
\label{bound}
\end{equation}
In the recent past where $\Omega_{\rm de}^{\rm eff}$ is not
negligible, this leads to a stringent constraint on the deviation
of the equation of state from -1. It should be noted that
although $\left \vert 1+w_{\rm eff} \right \vert \Omega_{\rm
de}^{\rm eff}$ is constructed simply out of the scale factor, $a$,
and its derivatives, neither $w_{\rm eff}(z)$ nor $\Omega_{\rm
de}^{\rm eff}(z)$ are uniquely defined as functions of redshift in
models such as these where the scalar field interacts with normal
matter.   As a result, it is possible to define $\Omega_{\rm
de}^{\rm eff}$ so that it vanishes and even becomes negative in
the past.  If such a definition is made, then one would (unless
the $\left \vert 1+w_{\rm eff} \right \vert \Omega_{\rm de}^{\rm
eff}$ also happens to  vanish) predict that $w_{\rm eff}$
diverges, and hence deviates significantly from $-1$. A behaviour
such as this was noted in Refs. \cite{tsu1,tsu2}. As a result, an
apparent effective deviation from $\Lambda$CDM can be deduced.
However, because of the freedom to redefine $\Omega_{\rm de}^{\rm
eff}$ and hence $w_{\rm eff}$, one should not rush to assign any
physical meaning to the divergence of $w_{\rm eff}$, and deduce
that it represents a significant deviation from  $\Lambda$CDM,
since one could always remove this divergence by choosing to
define $\Omega_{\rm de}^{\rm eff}$ it such a way that it is
positive definite. In all cases, the bound (\ref{bound})
gives an intrinsic measure of the deviation of the background
cosmology from $\Lambda$CDM, and it all cases it is small.
Therefore, the predicted late time cosmology is observationally
very close to $\Lambda$CDM. Additionally, the prospects for being
able to detect such small deviations for $\Lambda$CDM at the
background level in the near future are poor. Of course, as we
have already mentioned, detectably at the perturbative level might
be within reach.

This said, the thin-shell constraints do not themselves rule out
larger deviations from $\Lambda$CDM.  It may be that $\beta
\phi/\mpl$ has undergone relatively large changes in the past i.e.
much larger than $\mathcal{O}(10^{-4})$, but that we now just
happen to live at a point in time when $\beta \Delta \phi_{bc}
/\mpl < 10^{-4}$.  This would, however, be a fairly remarkable
coincidence and would inevitably require a great deal of fine-tuning of
the theory and the initial conditions. To avoid this
new coincidence problem,  we would have to require that the
cosmological changes in $\beta\phi/\mpl$ have been smaller than
$\mathcal{O}(10^{-4})$ in the recent past which would in turn, as
we illustrated above, constrain any deviations from $\Lambda$CDM
to be unobservably small. We note, however, that deviations
can be expected on very small scales, as in the original chameleon
model \cite{green}.

In this section we have sketched how the thin-shell requirement for
laboratory test masses place a very strong
constraint on the recent cosmological evolution of $\phi$, and
generally constrains any deviations from $\Lambda$CDM in the
predicted cosmology to be small.  This is not, however, a `water
tight' constraint as it may be possible to circumvent it by
requiring a seemingly improbable cosmological evolution wherein
such bodies would only have developed thin-shells in the recent
(in the cosmological sense) past.   The laboratory constraints
which we will derive in what follows cannot be avoided in this
way.

\section{Inverse Square Law Constraints}
In the weak field limit, the gravitational force due to a small
body drops off as $1/r^2$, where $r$ is the distance to the body's
centre of mass.  If there is an additional scalar degree of
freedom to gravity with constant mass $m_{\phi}$, the force
instead drops off as:
$$
\frac{\left(1+ \alpha(1+m_{\phi}r) e^{-m_{\phi}r})\right)}{r^2},
$$
where $\alpha$ parametrizes the strength with which the scalar degree of freedom couples to matter.  In $f(R)$ theories $\alpha = 2\beta^2 = 1/3$.   When $m_{\phi}r \ll 1$ or $m_{\phi}r \gg 1$, the force still drops off, approximately, as $1/r^2$, however there would be a noticeable deviation from this behaviour over scales $r \sim 1/m_{\phi}$.  If, as in chameleon theories, $m_{\phi}$ is not a constant but instead undergoes $\mathcal{O}(1)$ or greater variations, the behaviour of the force is more complicated but generally not of inverse square law form.

It is often assumed that what  is needed for an $f(R)$ theory to
avoid the constraints of inverse square law tests, is that the
test bodies develop thin-shells.  Generally, however, this is not
the case.  The presence of a thin-shell causes the chameleonic
force due to a  body to drop off much faster than $1/r^2$ near the
surface of the body.  Far from the body, the force has a
Yukawa form, although as a result of the fast drop-off near the
surface, it is much smaller than one would normally expect it to
be.   If two thin-shelled bodies are sufficiently close however
then they would be inside the region where the faster drop-off is
occurring. In these cases the detectable violation of the inverse
square law can be much larger than one might expect.

A number of different experiments have searched for violations
of the inverse square law.  For gravitational strength forces, i.e.
$\alpha \sim \mathcal{O}(1)$, the best constraints are currently
provided by the E\"{o}t-Wash experiment \cite{EotWash}.

The E\"{o}t-Wash experiment \cite{EotWash} consists of two plates:
the attractor and the detector. The detector is $0.997\,{\rm mm}$
thick and made out of molybdenum. The detector has 42 $4.767\,{\rm
mm}$ diameter holes bored into it in a pattern with $21$-fold
azimuthal symmetry.  The attractor is similar and consists of a
$0.997\,{\rm mm}$ thick molybdenum plate with 42 $3.178\,{\rm mm}$
diameter, arranged in a pattern with $21$-fold azimuthal symmetry,
mounted on a thicker tantalum disc with 42 holes, each with
diameter $6.352\,{\rm mm}$.  The holes in the lower tantalum ring
are displaced so that the torque on the detector due to the
attractor from forces, such as Newtonian gravity, that have a
$1/r^2$ behaviour vanishes.  The detection of a non-zero torque
would therefore indicate the presence of either a correction to
gravity with a behaviour different from $1/r^2$ or the presence of
a new force that also did not behave as $1/r^2$.

\subsection{Chameleonic Force \& Torque}
We now calculate the force, due to a chameleonic scalar field,
$\phi$, on one plate due to the other lying parallel to it.  From
this we calculate the chameleonic contribution to the torque.

In a background, where $\phi = \phi_b$ far from the plates, the
chameleonic force per unit area between two parallel plates, of
the same or similar compositions, both with thin-shells and with a
distance of separation $d$ between their two facing surfaces was
found, under certain conditions, in Ref. \cite{chamstrong,
chamcas}.  In Appendix \ref{appA} we generalise those formulae. We
find that the chameleonic force between two parallel circular
plates, with radius $r_p$ and thickness $D \ll r_p$, and
separation $d \ll r_p$ is given by:
\begin{eqnarray}
\frac{F_{\phi}}{A} = &&V(\phi_0) - V(\phi_{b}) -
V_{b}^{\prime}(\phi_0 - \phi_{b}) \label{eqn1} \\ &&+
\frac{(V_{b}^{\prime} - V_{c}^{\prime})^2}{2m_b^2 C(m_b r_p)}
E(\phi_c,\phi_b,m_br_p). \nonumber
\end{eqnarray}
where $\phi_0(d)$ is defined to the values of $\phi$ midway
between the two plates, and formulae for it are provided in Ref.
\cite{chamcas}.  We have also defined:
\begin{eqnarray}
E &=& 1 + 2C(m_b r_p)D(\phi_c,\phi_b) - \sqrt{1+4C D}, \\
C &=& \frac{(e^{m_b r_p} - 1/2)}{(e^{m_b r_p} -1)^2}. \\
D &=& \frac{m_{b}^2\left[V(\phi_b) - V(\phi_c) - V_{,\phi}(\phi_c)(\phi_b - \phi_c)\right]}{(V_{,\phi}(\phi_b) - V_{,\phi}(\phi_c))^2}.
\end{eqnarray}
The last term in Eq. (\ref{eqn1}) represents the only difference
between the generalized force formula and the one presented in
Refs. (\cite{chamstrong}) \& (\cite{chamcas}), and we note that
the extra term is independent of the separation $d$.  When $C(m_b
r_p) D(\phi_c, \phi_b) \ll 1$, the last term in Eq. (\ref{eqn1})
is negligible.  We note that $C(m_b r_p) \ll 1$ when $m_b r_p \gg
1$, and so whenever $m_b r_p \gg 1$, the last term is always
negligible.

The details of how $F_{\phi}(d)/A$ drops off with $d$ will
depend on the form of $V(\phi)$.
For many choices of $V(\phi)$, e.g. $V(\phi) \propto
\phi^{n}$ for $n < -2$ or $n > 2$, one finds that $F_{\phi}(d)/A$
drops off faster than $1/d$, for all $d > d_{2}$ is small compared
to both $r_{p}$ and the radius, $r_{h}$, of holes in the plates.
Indeed, this will certainly be the case, no matter what form
$V(\phi)$ takes, if $m_b r_{h} \gg 1$ where $m_b =
m_{\phi}(\phi_b)$.   Provided this is the case, we can define the
potential energy, $\mathcal{V}_{\phi}(d)$ due to the chameleonic
force for two plates with separation $d \ll R_{h}$ thus:
\begin{equation}
\mathcal{V}_{\phi}(d) \approx A \int_{d}^{\infty} \left(\frac{F_{\phi}(s)}{A}\right)\,\dd s.
\end{equation}
The faster $1/d$ drop off has been used to set the upper limit of the above integral to $\infty$.

In the E\"{o}t-Wash experiment  the plates have a number of holes
in them.  This means that as one plate is rotated, by an angle
$\theta$ say, relative to another, the surface area, $A(\phi)$,
of one plate that faces the other changes.  Note that $F_{\phi}/A$
does not depend on $A$. The torque due to the chameleonic force is
given by the rate of change of the potential ${\mathcal
V}_{\phi}(d)$ with $\theta$:
\begin{equation}
T_{\phi}(d) = \frac{\dd A}{\dd \theta} \int_{d}^{\infty} \left(\frac{F_{\phi}(s)}{A}\right)\,\dd s.
\end{equation}
We therefore have:
\begin{equation}
T_{\phi}(d) = a_{T} \int_{d}^{\infty} \left(\frac{F_{\phi}(s)}{A}\right)\,\dd s, \label{Teqn1}
\end{equation}
where $a_{T}  = \dd A / \dd \theta$ is a constant that depends only on the details of the experimental set-up rather than the theory being tested. For the 2006 E\"{o}t-Wash experiment \cite{EotWash} we find
$$a_{T} = 3.0 \times 10^{-3}\,{\rm m}^2.$$

If $F_{\phi}/A$ drops off too slowly over scales of the order of
$r_{h}$ then a more complicated analysis must be performed, and
knowing the force between two infinite parallel plates is no
longer enough to find a good approximation to the torque. Instead
a full numerical analysis would have to be undertaken to get
accurate results. This said, for $d \gtrsim r_{h}$, we do not
expect $F_{\phi}(d)$ to depend strongly on $\theta$ because the
effect of the holes will be largely smeared out over separation
distances much larger $r_{\rm h}$.  On scales $\ll r_{h}$, we
found that $F_{\phi} \propto A(\phi)$.  Since $T_{\phi} = \dd
V_{\phi}/\dd \theta$, $\dd V_{\phi}/\dd d = F_{\phi}$, and we
expect $F_{\phi}$ to be largely $\theta$ independent for $d \gg
r_{h}$ and $\propto A(\theta)$ on smaller scales, we expect, to
within an order of magnitude, that:
\begin{equation}
T_{\phi}(d) \approx  a_{T} \int_{d}^{r_{\rm h}}
\left(\frac{F_{\phi}(s)}{A}\right)\,\dd s. \nonumber
\end{equation}
in these cases, where once again $a_{T} = \dd A / \dd \theta$.  By
picking $r_{\rm h}$ as an upper bound for the integral we are
probably under estimating the torque as we are dropping the
contributions from larger separations.

\subsection{The Effect of an Electrostatic Shield}
Up to now we have not considered the role played by the
electrostatic shield.  Because the shield is so thin ($d_{\rm s} =
10\,\mu{\rm m}$) compared to the plates but has similar density to
the plates, it is safe to say that the shield will only have a
thin-shell when the plates have thin-shells.   Assuming the plates
do have thin-shells, we define $m_{\rm s}$ to be the mass the
chameleon would have deep inside the shield if the shield has a
thin-shell i.e. $m_{\rm s} = m_{\phi}(\phi_{\rm s})$ where
$V^{\prime}(\phi_{\rm s}) = -\beta\rho_{\rm s}/\mpl$. Since the
shield is sandwiched between the two plates, the thin-shell
condition for the shield is simply $m_{\rm s} d_{\rm s} \gtrsim
1$.  When the shield has a thin-shell, its presence attenuates the
chameleonic force and torque on the detector due to the attractor
by a factor of $\exp(-m_{\rm s}d_{\rm s})$.  Since $\exp(-m_{\rm
s}d_{\rm s}) \approx 1$ in the absence of a thin-shelled shield,
we can take account of the shield, thin-shell or not, by changing
the definition of $T_{\phi}$ thus:
\begin{equation}
T_{\phi}(d) \approx a_{T} e^{-m_{\rm s} d_{\rm s}}\int_{d}^{R_{h}} \left(\frac{F_{\phi}(s)}{A}\right)\,\dd s. \label{Teqn}
\end{equation}
This expression provides a very good approximation for theories in
which the precise value of $r_{h}$ is unimportant (e.g. ones with
$m_b r_{h} \gg 1$) and an order of magnitude estimate otherwise.

\subsection{Inverse Square Law Constraints}
The 2006 E\"{o}t-Wash experiments requires that
\begin{eqnarray}
\vert T_{\phi}(d=55\,\mu{\rm m}) \vert &<& 0.87 \times 10^{-17}\,{\rm Nm},\nonumber
\end{eqnarray}
with $95\%$ confidence.   We define $T_{\phi}(d = 55 \mu\,{\rm m}) = a_{T}\Lambda^3_{T}$ and find that the above bounds correspond to:
\begin{eqnarray}
\Lambda_{T} < 0.89 \times 10^{-12}\,{\rm GeV},
\end{eqnarray}
Importantly this is smaller than the energy scale associated with dark energy: $\Lambda_{\rm de} = 2.4 \times 10^{-12}\,{\rm GeV}$;  $\rho_{\rm de} = \Lambda_{\rm de}^{4}$.

Using our expression Eq.(\ref{Teqn}) for the chameleonic torque,
we find that the constraints we must apply are as follows:
\begin{eqnarray}
e^{-m_{\rm s} d_{\rm s}}\int_{55 \, \mu{\rm m}}^{R_{h}} \left(\frac{F_{\phi}(s)}{A}\right)\,\dd s &<& 7.0 \times 10^{-37}\,{\rm GeV}^3. \label{Eotcons1}
\end{eqnarray}

\section{Application of Constraints to $f(R)$ theories}
\subsection{Chameleonic force}

The chameleonic force per unit area between two parallel plates
is given by Eq. (\ref{eqn1}).  To prevent large
deviations from general relativity occurring over solar system,
and smaller, scales, one must require that $f(R) \approx R + h(R)$
where $\vert h^{\prime}(R)\vert \ll 1$ and $\vert h(R)/R \vert \ll
1$.  In this case the expression for $F_{\phi}/A$ becomes:
$$
\frac{F_{\phi}}{A} \approx \frac{\mpl^2}{2}\left[\left(R_0-R_b\right)h^{\prime}(R_0)+ \left(h(R_b)-h(R_0)\right) + \mathcal{F}_0\right].
$$
where for $R_c \gg R_b$
\begin{eqnarray}
\mathcal{F}_0 &=& \frac{R_c^2 h^{\prime \prime}(R_{b})}{4 C(m_b r_{\rm p})} \mathcal{E}_{0}, \\
\mathcal{E}_0 &=& 1+2C(m_b r_{\rm p})\mathcal{D}_{0}(R_c, R_b) - \sqrt{1+4C\mathcal{D}_0}, \label{calE}\\
\mathcal{D}_0 &=& \frac{h(R_c) - h(R_b) - (R_c-R_b) h^{\prime}(R_b)}{h^{\prime \prime}(R_b) R_c^2}.
\end{eqnarray}
where $r_{\rm p}$ is the radius of the parallel plate(s).
We shall now consider several potential forms for $h(R)$.

\subsection{Logarithmic potentials}
We begin  by considering a simple chameleon  gravity model that
was recently suggested in Ref. \cite{kaloper} for a general
$\beta$. The theory, when written as a chameleon theory, would
have a potential $V(\phi) = V_{0} - \Lambda_0^4 \ln (\phi/\mpl)$,
it was suggested that this would result in an experimentally
viable and cosmologically interesting dark energy model, where
$\Lambda_0^4 /\mpl^2 \sim \mathcal{O}(H_0^2)$ \cite{kaloper} for
$\beta\le 1/4\sqrt{3}$. We will analyse the same model in the
$f(R)$ setting where $\beta=1/\sqrt{6}> 1/4\sqrt 3$ and show that
local tests already lead to difficulties, see also \cite{kaloper2}.

On laboratory scales we would have  $f(R) \approx 1+h(R)$ and so we find:
$$
h(R) = - \frac{2\Lambda_0^4}{\mpl^2} \left[ \frac{V_0}{\Lambda_0^4} + \ln(2\beta) + \ln\left(\frac{\mpl^2 R}{2\Lambda_0^4}\right) + 1\right].
$$
Assuming that $m_b \ll m_c$ where $m_c$ is the chameleon mass inside the plates, it follows that $F_{\phi}/A$ has the following form:
\begin{equation}
\frac{F_{\phi}}{A} = \Lambda_0^4\left[ \ln\left(\frac{R_0}{R_b}\right) + \frac{R_b}{R_0} - 1 \right] +\frac{1}{2}\Lambda_0^4 \frac{m_{c}^2 \mathcal{E}_{0}}{m_b^2 C(m_b r_{\rm  p})}. \label{Flneqn}
\end{equation}
where
$$
\mathcal{E}_{0} = 1 + \frac{2C(m_b r_{\rm p}) m_{b}}{m_c} - \sqrt{1+ \frac{4C(m_b r_{\rm p})m_b}{m_c}}.
$$
When $C(m_b r_{\rm p}) m_b / m_c \ll 1$ we have $\mathcal{E}_{0} \approx 2C^2 m_b^2/m_c^2$ and so the last term in Eq. (\ref{Flneqn}) is:
$$
\Lambda_0^4\frac{m_{c}^2}{2m_b^2 C(m_b r_{\rm  p})}\mathcal{E}_{0} \approx \Lambda_0^4C(m_b r_{\rm p}).
$$
Alternatively if $C(m_b R)m_b / m_c \gg 1$ we would have:
$$
\Lambda_0^4\frac{m_{c}^2}{2m_b^2 C(m_b r_{\rm  p})}\mathcal{E}_{0} \approx \Lambda_0^4 \frac{m_{c}}{m_{p}}.
$$
If $C(m_b r_{\rm p}) \ll 1$ i.e. $m_b r_{\rm p} \gg 1$, then it is clear that this last term is always small compared to the other terms, however if $m_b r_{\rm p} \ll 1$, then the last term will dominate the expression for the force.

The chameleon mass for a given $R$ in this set-up is
$$m_{\phi}(R) = \frac{\mpl R}{\sqrt{6}\Lambda_0^2}.$$
In between the two plates, $\phi$ satisfies \cite{chamcas}:
$$
\frac{\dd^2 \phi}{\dd z^2} =  V_{,\phi}(\phi) - V_{,\phi}(\phi_b),
$$
and $\phi_{0}$ is defined to be value of $\phi$ midway between the two plates (i.e. a distance $d/2$ from either plate), where by symmetry $\dd \phi/\dd z  =0$. Integrating the above equation we therefore have:
$$
\left(\frac{\dd \phi}{\dd z}\right)^2 = 2(V(\phi) - V(\phi_0) - V_{,\phi}(\phi_b)(\phi-\phi_0)).
$$
Integrating this again and defining $\phi_{\rm s} \sim
\mathcal{O}(\phi_c)$ to be the value of $\phi$ on the surface of
the plates, we have:
$$
\frac{d}{\sqrt{2}} = \int_{\phi_{\rm s}}^{\phi_{0}} \frac{\dd x}{\sqrt{V(x)-V(\phi_0) - V_{,\phi}(\phi_b)(\phi-\phi_0)}}.
$$
Following Ref. \cite{chamcas}, when $m_{b} \ll m_{0} \ll m_{c}$ we
have that $\phi_{s} \sim \mathcal{O}(\phi_c) \ll \phi_{0} \ll
\phi_b$ and so $V(\phi)-V(\phi_0) - V_{,\phi}(\phi_b)(\phi-\phi_0)
\approx V(\phi) -V(\phi_0) = \Lambda_0^4\ln(\phi_0/\phi)$ and so:
$$
\frac{d}{\sqrt{2}}  \approx \frac{\phi_{0}}{\Lambda_0^2} \int_{0}^{1} \frac{\dd x}{\sqrt{\ln(1/x)}}.
$$
Noting that $m_{0}=\Lambda_{0}^2/\phi_0$ and evaluating the integral we find:
$$
m_{0} d = \frac{\mpl R_{0} d}{\sqrt{6}\Lambda_0^2} = \sqrt{2\pi}.
$$
Now $m_{b} \ll m_{0} \ll m_{c}$ is clearly equivalent to $R_c \gg R_0 \gg R_b$, and in these cases we therefore have:
\begin{equation}
R_{0} = \frac{2\sqrt{3\pi} \Lambda_0^2}{\mpl d}. \label{m0eqn}
\end{equation}
It follows that, irrespective of the value of $\mathcal{E}_{0}$, $F_{\phi}/A$ drops off more slowly than $1/d$ for all $m_b d \ll 1$. In the latest version of the E\"{o}t-Wash experiment \cite{EotWash}, the plate radius, $r_{\rm p}$,
is $3.5\,{\rm cm}$, and the smallest hole radius is $1.6\,{\rm mm}$. The pressure of the laboratory vacuum is $10^{-6}\,{\rm torr}$ which corresponds to a background density of $\rho_b = 1.6 \times 10^{-9}\,{\rm kg\,m}^{-3} = 6.7 \times 10^{-30}\,{\rm GeV}^4$.

Now if the vacuum region is large enough then $R_b =
\rho_{b}/\mpl^2$ and so $m_{b} =  \bar{m}_{b} \equiv
\rho_{b}/\sqrt{6}\Lambda_0^2 \mpl$.  However, it was shown in
Refs. \cite{chamstrong, chamPVLAS, chamcas} that if the vacuum
region only have length scale $L_{\rm vac}$ and $\bar{m}_b \ll
1/L_{\rm vac}$, then generically $m_b \sim \mathcal{O}(1/L_{\rm
vac})$.  For the moment we only assume that $r_{\rm p}/L_{\rm vac}
\ll 1$.

We therefore find that for $d=55\,\mu{\rm m}$ we  have $m_{b}d <
1$ for all $\Lambda_{0} > 5.6 \times 10^{-19}\,{\rm GeV}$, $m_{b}
r_{\rm h} < 1$ for $\Lambda_{0} > 3.0 \times 10^{-18}\,{\rm GeV}$
and $m_b r_{\rm p} < 1$ for $\Lambda_{0} > 1.4 \times
10^{-17}\,{\rm GeV}$.  The suppression factor due to the
electrostatic shield is $\exp(-m_{s}d_{\rm s})$ where:
$$
m_{\rm s}d_{\rm s} = \frac{\beta \rho_{\rm shield} d_{\rm s}}{\mpl \Lambda_{0}^2} = 0.30 \left(\frac{10^{-12}\,{\rm GeV}}{\Lambda_0}\right)^2,
$$
where we have used $d_{\rm s} = 10\,\mu{\rm m}$ and $\rho_{\rm s}
= 8.3\,{\rm g\,cm}^{-3}$.   Thus, whenever $m_{\rm s} d_{\rm s}
\lesssim 1$, we are therefore firmly in the $m_b r_{\rm p}\,
m_{b}r_{\rm h}\, \ll 1$ region and hence $C(m_{b}r_{\rm p}) \gg
1$.  Thus $C(m_{b}r_{\rm p}) \approx 1/2m_{b}^2r_{\rm p}^2$ and:
\begin{eqnarray}
\frac{F_{\phi}}{A} \approx &&\Lambda_{0}^4\left[-\ln(m_{b}d/\sqrt{2\pi}) + m_{b}d/\sqrt{2\pi}-1\right] \nonumber \\ &&+ \Lambda_0^4 m_{c}^2 r_{\rm p}^2\mathcal{E}_{0}\nonumber
\end{eqnarray}

From Eq.(\ref{Teqn}), in the absence of the electrostatic shield,
the chameleonic torque for $r_{h} \gg d$ would be:
\begin{eqnarray}
T_{\phi} \approx &&a_{T} \Lambda_{0}^4 r_{h}\left[\ln\left(\frac{\sqrt{2\pi}}{m_{b}r_{h}}\right) + \frac{m_{b}r_{h}}{\sqrt{2\pi}} \right. \\ &&\left.+  m_{c}^2 r_{\rm p}^2 \mathcal{E}_{0}\right].\nonumber
\end{eqnarray}
Here $m_{c}$ is the chameleon mass inside the plates which have density $\rho_{c} \approx 10.2 \,{\rm g\,cm}^{-3}$.

We note that the requirement that the plates have a thin-shell
constrains the value of $m_{b}$, and it is important to check that
this constraint holds.  Conservatively, the thin-shell constraints
for the plate require $\beta(\phi_b - \phi_c)/\mpl < \Phi_{N}/3$
where $\Phi_{N}$ is the Newtonian gravitational potential of the
whole experiment at the surface of the plate.  Since the geometry
of the experiment is complicated, we do not calculate $\Phi_{N}$.
Instead, we estimate $\Phi_{N}/3 \lesssim 10^{-26}$, and so
$\phi_b - \phi_c \lesssim   7\times 10^{-8}\,{\rm GeV}$. Given
that $\rho_c \gg \rho_b$, we take $\phi_c \ll \phi_b$ and then
from $m_b = \Lambda_0^2/\phi_b$ we must have
\begin{equation}
1/m_{b} < 14\,{\rm m} \left(\frac{10^{-12}\,{\rm GeV}}{\Lambda_0}\right)^2. \label{thincondvac}
\end{equation}
If this condition does not hold, then the plates would not have
thin-shells and the E\"{o}t-Wash data would automatically rule out
the theory.  The experiment takes place inside a vacuum chamber
with smallest dimension $L_{\rm vac}=0.2\,{\rm m}$
\cite{Kapnerpriv}.  We assume that the walls of the vacuum chamber
have thin-shells. Approximating the walls of the vacuum chamber
perpendicular to the shortest dimension as being parallel plates,
we use Eq. (\ref{m0eqn}) above to tell us that when the background
density of matter in the vacuum chamber is very small, we have in
the centre of the chamber:
$$
R_{b} = \bar{R}_{\rm vac} = \frac{2\sqrt{3\pi} \Lambda_0^2}{\mpl L_{\rm vac}}.
$$
This formula holds as long as $\bar{R}_{b} = \rho_{b}/\mpl^2
\lesssim \bar{R}_{\rm vac}$.  In the opposite limit we just have
$R_b = \bar{R}_b$.  In all cases we have $R_b \geq \bar{R}_{\rm
vac}$ and so:
\begin{equation}
m_{b} \geq \frac{\mpl \bar{R}_{\rm vac}}{\sqrt{6}\Lambda_0^2} = \frac{\sqrt{2\pi}}{L_{\rm vac}} \approx 12\,m^{-1} \label{mbeqn}
\end{equation}
and so condition (\ref{thincondvac}) is always satisfied for
$\Lambda_0 \lesssim 1.2\times 10^{-11}\,{\rm GeV}$.

Given Eq. (\ref{mbeqn}), we find that for the allowed values of $\Lambda_0$ we always have:
$$
\frac{C(m_{b} r_{\rm p})m_{b}}{m_{c}} \approx \frac{1}{2m_{b} m_{c} r_{\rm p}^2} \ll 1,
$$
where we have used $m_{c} = \rho_{c}/\sqrt{6}\mpl \Lambda_0^2$.  Thus:
$$
\mathcal{E}_{0} \approx 2C^2 m_{b}^2/m_c^2 = (m_{c}m_b r_{\rm p})^{-4} /2.
$$
It follows that, in the absence of the electrostatic shield, the chameleonic torque for $r_{h} \gg d$ is
\begin{eqnarray}
T_{\phi} \approx &&a_{T} \Lambda_{0}^4 r_{h}\left[\ln\left(\sqrt{2\pi}{m_{b}r_{h}}\right) + \frac{m_{b}r_{h}}{\sqrt{2\pi}} \right. \\ &&\left.+  \frac{1}{2 m_{b}^2 r_{\rm p}^2}\right].\nonumber
\end{eqnarray}
Including the suppression factor due to the electrostatic shield, which  is $\exp(-m_{s}d_{\rm s})$, we therefore find the following constraint on $y_0=\Lambda_{0}/(10^{-12}\,{\rm GeV})$:
\begin{equation}
e^{-0.30 y_{0}^{-2}/4} y_{0} < 0.21, \nonumber
\end{equation}
which gives $y_0 < 0.37$ and so:
\begin{equation}
\Lambda_{0} < 3.7 \times 10^{-13}\,{\rm GeV}.
\end{equation}

Cosmologically, the mass of the scalar field at the minimum of its
 potential is given by
$$
m_{\rm cos} = \left[\sqrt{\frac{3}{2}} \frac{\Omega_m M_{pl} H}{\Lambda_0^2}\right] H,
$$
and the value of $\phi$ at this minimum is given by:
$$
\frac{\beta \phi_{\rm cos}^{\rm min}}{\mpl} = \frac{\Lambda_{0}^4}{3\Omega_m H^2 \mpl^2}.
$$
Now $H = 2.1 h \times 10^{-42}\,{\rm GeV}$ and from WMAP \cite{wmap}: $\Omega_{m} =0.127h^{-2}$ and $h=0.73$.
We find that
\begin{eqnarray}
m_{\rm cos}/H = \frac{1.1}{y_{0}^2},\\
\frac{\beta \phi_{\rm cos}^{\rm min}}{\mpl}  = 0.06 y_{0}^4
\end{eqnarray}
and so the E\"{o}t-Wash constraint on $\Lambda_0$ gives:
$$
m_{\rm cos}/ H > 8.
$$
At its minimum then, the $\phi$-field is still heavy today.
This should be contrasted with the requirement obtained in
\cite{kaloper} that the mass of the $\phi$-field should be small
compared to the Hubble rate in order to drive acceleration. Here
we find that local tests and the thin shell requirement impose
that the mass of the $\phi$-field at the cosmological minimum is
so large that the field must sit there on cosmological
scales. It is easily checked that $m_{\rm cos}/H$ is a
decreasing function of time, and so in the past $\phi$ was heavier
still relative to $H$. Therefore $\phi$ will have remained stuck
close to the minimum of its effective evolution throughout the
matter era. Additionally the E\"{o}t-Wash constraint on
$\Lambda_0$ implies that:
$$
\frac{\beta \phi_{\rm cos}^{\rm min}}{\mpl} < 0.001
$$

We have considered the potential $V(\phi) = V_{0} -
\Lambda_0^4\ln(\phi/\mpl)$.  The constraint on $\Lambda_0$ implies
that $\Lambda_{0}^4/3\mpl^2 H^2 < 0.00026$ and so if $V(\phi)$ is
to be the source of dark energy and there is to be a realistic
amount of it today, we would need  $V_{0} \gtrsim
\mathcal{O}(1000) \Lambda_{0}^4$. Note that this is very different
from the original scenario envisaged in ref. \cite{kaloper}, where
$V_{0} \sim \Lambda_{0}^4$ so that the whole potential could be
written in the form $V = -\Lambda_0^4 \ln (\phi/M)$ where $M \sim
\mathcal{O}(\mpl)$.   The E\"{o}t-Wash constraint on $\Lambda_0$
therefore rules out a scenario where $V_{0} \sim  \Lambda_{0}^4$
for $\beta=1/\sqrt{6}$, confirming the cosmological obstruction
noted when $\beta>1/4\sqrt 3$. If we moved away from $f(R)$
theories and allowed for different couplings, we would find
similar constraints on $\Lambda_{0}$ for other $\mathcal{O}(1)$
values of the coupling $\beta$.

Relaxing the constraint $V_{0} \sim \Lambda_{0}^4$ and allowing
much smaller values of $\Lambda_0$, it should also be noted that
the conservative thin-shell constraint for a test mass with
density $\gg \Lambda_0^4$ on the cosmological value of $\phi$ (as
derived in Section \ref{sec:cosmo}) actually provides a stronger
constraint on the cosmological value of the field today and as
such gives a tighter bound on $\Lambda_0$. Specifically Eq.
(\ref{phicoscon}) implies:
\begin{equation}
\frac{\beta \phi_{\rm cos}}{\mpl} < 10^{-4} \Leftrightarrow \Lambda_0 < 1.8 \times 10^{-13}\,{\rm GeV}.
\end{equation}
This leads to the following constraint on the mass of $\phi$ at its minimum cosmologically:
\begin{equation}
m_{\rm cos}/H > 35 \label{pert}.
\end{equation}
Since $m_{\rm cos}/H \gg 1$, $\phi$ lies close to its cosmological
minimum and so, in the Jordan frame, by Eq. (\ref{PhiJordan}):
$$
-2\Phi^3 V_{,\Phi} = \frac{\mpl}{\beta} \Phi^2 V_{,\phi} = -\rho_{\rm matter},
$$
where $\Phi = e^{-2\beta \phi/\mpl}$ and so $\Phi \approx 1$.  To leading order with $p = \ln a$
we have $\phi_{p} \approx 3\phi$. Therefore
$$
\frac{\Phi_{p}}{\Phi} = -2\frac{\beta \phi_{p}}{\mpl} \approx \frac{-6\beta \phi}{\mpl} = -\frac{6\Lambda_0^4}{\rho_{\rm matter}},
$$
and to the same order
$$
\frac{\Phi_{pp}}{\Phi} \approx -\frac{18 \beta \phi}{\mpl} = -\frac{18 \Lambda_0^4}{\rho_{\rm matter}}.
$$
We define $\theta = \Lambda_0^4/\rho_{\rm matter}$ and then using
Eq. (\ref{weffeqn}) we arrive at
\begin{eqnarray}
(1+ w_{\rm eff}) \Omega_{\rm de}^{\rm eff} \approx &&
\frac{(2\Omega_{\rm de}^{\rm eff} - \Omega_{\rm m}^{\rm eff} + 2)\theta}
{3\theta-1}  \\ &&+ \frac{(\theta - \theta_0) \Omega_{\rm m}^{\rm
eff}}{1-3\theta} \nonumber
\end{eqnarray}
where $\theta_0$ is the value of $\theta$ at the present time.
Assuming that the Universe is flat ($k=0$) and taking $\Omega_{\rm
de}^{\rm eff}  = 0.76$, we find today when $t=t_0$:
$$
1+ w_{\rm eff}(t_0) \approx - 4.32 \theta_0 .
$$
Notice that the effective equation of state is below -1, this is a
consequence of the scalar-tensor character of the chameleon model.

The E\"{o}t-Wash constraint on $\Lambda_0$ gives
$$
\vert 1 + w_{\rm eff} \vert < 0.0085,
$$
while the thin-shell constraint on $\Lambda_0$ gives
$$
\vert 1 + w_{\rm eff} \vert < 10^{-4},
$$
which is in line with our expectations from Section \ref{sec:cosmo}.
Whilst the thin-shell constraint on the cosmology is much stronger
than the E\"{o}t-Wash bound, the cosmological constraint makes a
number assumptions above the nature of inhomogeneities in the
Universe, in particular about their scale at the present time. One
could presumably argue that the cosmological constraint
could be relaxed.  The same line of argument cannot be used for the
E\"{o}t-Wash constraint, and as such represents a strong
constraint on the magnitude of deviations from $\Lambda$CDM in
this model. As such, the model
cannot be distinguished from a $\Lambda$CDM model at the
background level. At the perturbative level, the situation is very
different as the bound (\ref{pert}) implies that density contrast
would have an anomalous growth on scales lower than $100 h^{-1}
{\rm Mpc}$. This may be testable in the near future with next generation
red-shift surveys\cite{spergel}.

The version of the logarithmic potential $f(R)$ theory suggested
in Ref. \cite{kaloper} required $\Lambda_{0}^4 \geq \rho_{\rm
matter}$ today i.e. $\Lambda_0 > 1.7 \times 10^{-12}\,{\rm GeV}$.
Even from a conservative point of view (and indeed for any $\beta$
where $2\beta^2 \sim \mathcal{O}(1)$), such a value of $\Lambda_0$
would produce a torque in the E\"{o}t-Wash experiment that is
almost $100$ times larger than the $95\%$-confidence level upper
bound. The scenario suggested in Ref. \cite{kaloper} is therefore
strongly ruled out by local tests of gravity.

\subsection{Power-law form}
In many cases \cite{Hu1,Hu2,Staro} one finds that for $R \gg H_0^2$,
where $H_0^2$ is the Hubble constant today,  $h(R)$ has a power
law form i.e.:
\begin{equation}
h(R) \approx
\frac{p}{p+1}\bar{R}\left(\frac{R}{\bar{R}}\right)^{p+1},
\end{equation}
for some $p \neq 0$ and some constant $\bar{R} > 0$. For a
chameleon mechanism to exist we need $V^{\prime} < 0$, $V^{\prime
\prime}>0$ and $V^{\prime \prime \prime} < 0$, and so must require
 $p < 1$.   Relative divergences from GR such as those
parametrized in the PPN formalism or measured by observing the
motions of planets would scale as $h(R)/R$ and $h^{\prime}(R)$ or
by the ratio of any variable component of the effective
cosmological constant to the local matter density .   However, the
E\"{o}t-Wash test probes changes in $V(\phi)$, which scales as
$h(R)$ and $R h^{\prime}(R)$, although they are only sensitive to
this when the chameleon mass  in the background, which scales as
$1/h^{\prime \prime}(R)$, is not too large.

In theories with $0 < p < 1$, both $h(R)/R$ and $V(\phi)$ would be largest for large values of $R$.  These theories would therefore diverge most markedly from General Relativity in the UV (i.e. large $R$) regime.   Increasing $\bar{R}$ would make both $h(R)/R$ and $h^{\prime \prime}(R)$ smaller, and so ultimately one could ensure compatibility with all laboratory tests by making $\bar{R}$ very large.  Provided $h^{\prime \prime}$ is not small, however, the changes in $V(\phi)$ that could be detected by the E\"{o}t-Wash experiment would increase.

If $-1 < p < 0$ then $h(R)/R$ and $h^{\prime}(R)$ are largest in
the IR regime where $R$ is small. However $V(\phi)$ still
increases with $R$,  and since $R$ increases as the separation of
the plates in the E\"{o}t-Wash experiment is decreased,  the
smaller the separations the stronger the potentially detectable
signal would be.  Ultimately compatibility with all local tests
could be ensured by making $\bar{R}$ small enough. Additionally in
all theories where $p > -1$,  $F_{\phi}/A$ would be dominated by
the $d$-dependent (i.e. $R_0$ dependent) terms and only weakly
depend on $R_b$ when $m_b d \ll 1$.

Finally, in theories with $p < -1$ both $V(\phi)$ and $h(R)/R$
would decrease with $R$.  This would mean that $F_{\phi}/A$ would
only very weakly depend on $d$ and generally be much smaller in a
given set-up than for the other classes of theories.   Again
compatibility with all local tests could be ensured by making
$\bar{R}$ small enough.

The $-1 < p < 0$ theories are the most testable type of theory as
they would result in deviations from GR in both the UV and IR
regimes.  In the UV regime there would be potentially detectable
fifth-forces between parallel plates, and in the IR regime the
ratio of the density dependent part of the effective cosmological
constant to the ambient matter density would increase
cosmologically at late times as the ambient density decreased.

In all of these theories:
\begin{eqnarray}
 \frac{F_{\phi}}{A} = &&\frac{M_{\rm Pl}^2 R_0 p^2 }{2(p+1)} \left(\frac{R_0}{\bar{R}}\right)^{p}  \label{FApower} \\ && + \frac{M_{\rm Pl}^2 R_b p}{2(p+1)} \left(\frac{R_b}{\bar{R}}\right)^{p}  - \frac{M_{\rm Pl}^2 R_b p}{2}  \left(\frac{R_0}{\bar{R}}\right)^{p} \nonumber \\ &&+ \frac{\mpl^2 p^2 R_{c}^2}{8 C(m_b r_{\rm p}) R_{b}}\left(\frac{R_b}{\bar{R}}\right)^{p}\mathcal{E}_{0} . \nonumber
\end{eqnarray}
where $\mathcal{E}_{0}$ is given in terms of $C$ and $\mathcal{D}_0$ by Eq. (\ref{calE}) and
$$
\mathcal{D}_{0} = \frac{1}{p(p+1)}\left[\left(\frac{R_b}{R_{c}}\right)^{1-p} + p\left(\frac{R_{b}}{R_{c}}\right)^{2} - (p+1)\left(\frac{R_b}{R_c}\right)\right].
$$
Note that the last term in Eq. (\ref{FApower}) is independent of the plate separation $d$ and vanishes in the limit $m_{b}r_{\rm p} \rightarrow \infty$.

\subsubsection{Relationship to chameleon theories}
Converting these theories to chameleon theories we have for $h^{\prime} \ll 1$:
$$
-2\frac{\beta \phi}{\mpl} = h^{\prime}(R) = p \left(\frac{R}{\bar{R}}\right)^{p},
$$
and so $R \propto \phi^{1/p}$ and $h(R),\,Rh^{\prime}(R) \propto \phi^{(p+1)/p}$. It follows that
$$
V(\phi) = {\rm const} + \frac{p^2 \mpl^2 \bar{R}}{2(p+1)} \left(\frac{-2 \beta \phi}{\mpl p}\right)^{\frac{p+1}{p}}.
$$
and so defining $n = -(p+1)/p$, we see that, neglecting the
constant term in the potential, $\vert V(\phi)\vert \propto
\vert\phi \vert^{-n}$.  In the context of chameleon theories these
potentials have been studied in great detail \cite{chamKA,
chamcos, chamstrong, chamPVLAS, chamcas}, and so we are able to
apply a raft of results to the analysis of these theories.

In appendix \ref{appB}, we show the mass of the chameleon field
for $m_{c} \gg m_{0} \gg m_{b}$ ($m_c$ is the chameleon mass deep
inside the plates and $m_b$ is the chameleon mass in the
background) is given by $m_0 d = a_p$ where:
\begin{eqnarray}
a_{p} &=& \sqrt{\frac{2}{1+p}} p^2 B\left(\frac{1}{2}, \frac{p}{(1+p)}\right) \qquad p \leq -1, \nonumber\\
a_{p} &=& \sqrt{\frac{2}{1+p}}B\left(\frac{1}{2},\frac{(1-p)}{2(1+p)}\right) \quad -1 \leq p \leq 1. \nonumber
\end{eqnarray}
Using $m_{0} \approx 1/3 h^{\prime \prime}(R_0)$ we therefore have when $R_{b} \ll R_0 \ll R_c$:
\begin{equation}
\frac{R_0}{\bar{R}} = \left(\frac{3p^2 a_{p}^2}{\bar{R}d^2}\right)^{\frac{1}{1-p}}.
\end{equation}
For what follows we also define:
\begin{eqnarray}
K_{p} = \left(3p^2 a_p^2\right)^{\frac{1+p}{1-p}}.\label{Kpeqn}
\end{eqnarray}
Using $m_{0} \approx 1/3 h^{\prime \prime}(R_0)$ we therefore have when $R_{b} \ll R_0 \ll R_c$:

Using the relationship between $R_{0}$ and $d$ derived above when
$R_{c} \gg R_{0} \gg R_{b}$ Eq. (\ref{FApower}) becomes:
\begin{eqnarray}
\frac{F_{\phi}}{A} &\approx& \frac{M_{\rm Pl}^2 K_{p} p^2\bar{R}}{2(p+1)} \left(\frac{1}{\bar{R}d^2}\right)^{\frac{1+p}{1-p}}G_{p}\left(\frac{m_b d}{a_p}\right) , \label{FApower1}\\
&+& \frac{\mpl^2 p^2 R_{c}^2}{8 C(m_b r_{\rm p}) R_{b}}\left(\frac{R_b}{\bar{R}}\right)^{p}\mathcal{E}_{0}. \nonumber
\end{eqnarray}
where
\begin{eqnarray}
G_{p}\left(\frac{m_b d}{a_p}\right) = &&1 + \frac{1}{p}\left(\frac{m_b d}{a_p}\right)^{\frac{2(1+p)}{1-p}} \\ &&- \frac{p+1}{p}\left(\frac{m_b d}{a_p}\right)^{\frac{2}{1-p}}.\nonumber
\end{eqnarray}
We note that $G_{p} = 0$ when $m_b d /a_p = 1$ which corresponds to $R_0 = R_b$. We now consider the integral:
\begin{equation}
I(d, r_{\rm h}) = \int_{d}^{r_{\rm h}} \frac{F_{\phi}(s)}{A} \dd s \nonumber
\end{equation}
The approximation used to calculate $R_0(d)$ breaks down when
$R_0 \approx R_b$, which corresponds to $m_b d / a_p \approx 1$. In
the case $m_b r_{\rm h}/a_p > 1$ we cannot simply use Eq.
(\ref{FApower1}) to calculate $I(d,r_{\rm h})$ as we must
integrate over values of $d$ for which Eq. (\ref{FApower1}) is not
valid. However we should be able to trust Eq. (\ref{FApower1}) for
smaller values of $d$.  For $m_b d/a_p \gg 1$ we expect an
exponential drop-off in the force, just as one would find in a
Yukawa theory at distances larger than the inverse mass of the
scalar field.  We therefore do not expect the dominant
contribution to $I(d,r_{\rm h})$ to come from values of $d <
a_p/m_{b}$. We also note that if $m_b r_{\rm h} /a_p \gg 1$ then
$m_b r_{\rm p} \gg 1$ as $r_{\rm h} < r_{\rm p}$ and as such the
second term in  Eq. (\ref{FApower1}) is negligible.  The first
term in Eq. (\ref{FApower1}) vanishes when $m_b d/a_p = 1$, and
since we do not expect a significant contribution to the integral
to come from larger separations, we evaluate $I(d,r_{\rm h})$ by
using Eq. (\ref{FApower1}) for $F_{\phi}(s)/A$ but if $m_b r_{\rm
h}/a_p$ we cut the integral off at a separation $a_p/m_{\rm h}$.
\begin{figure}[htb!]
\includegraphics*[width=6.5cm]{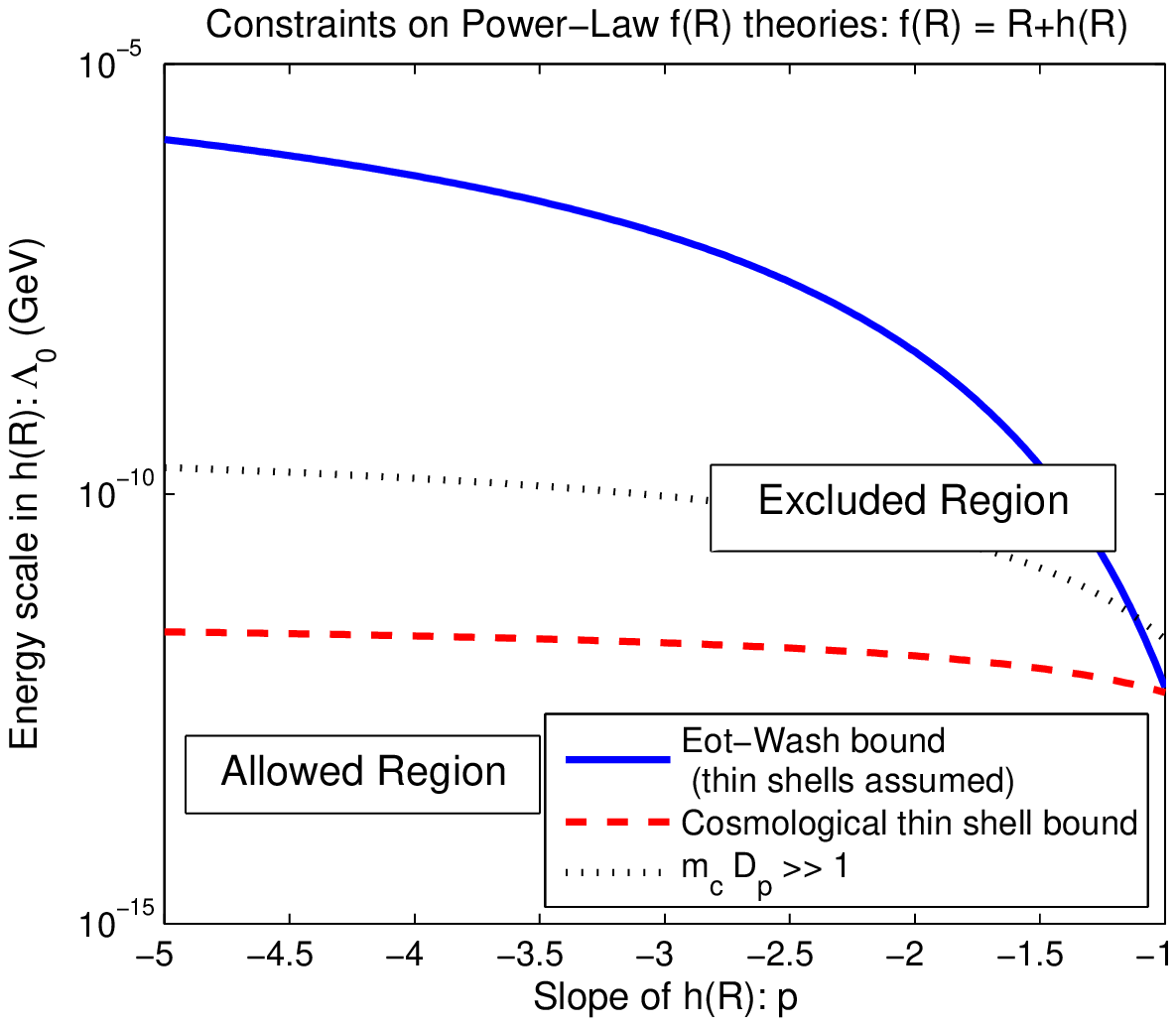}
\includegraphics*[width=6.5cm]{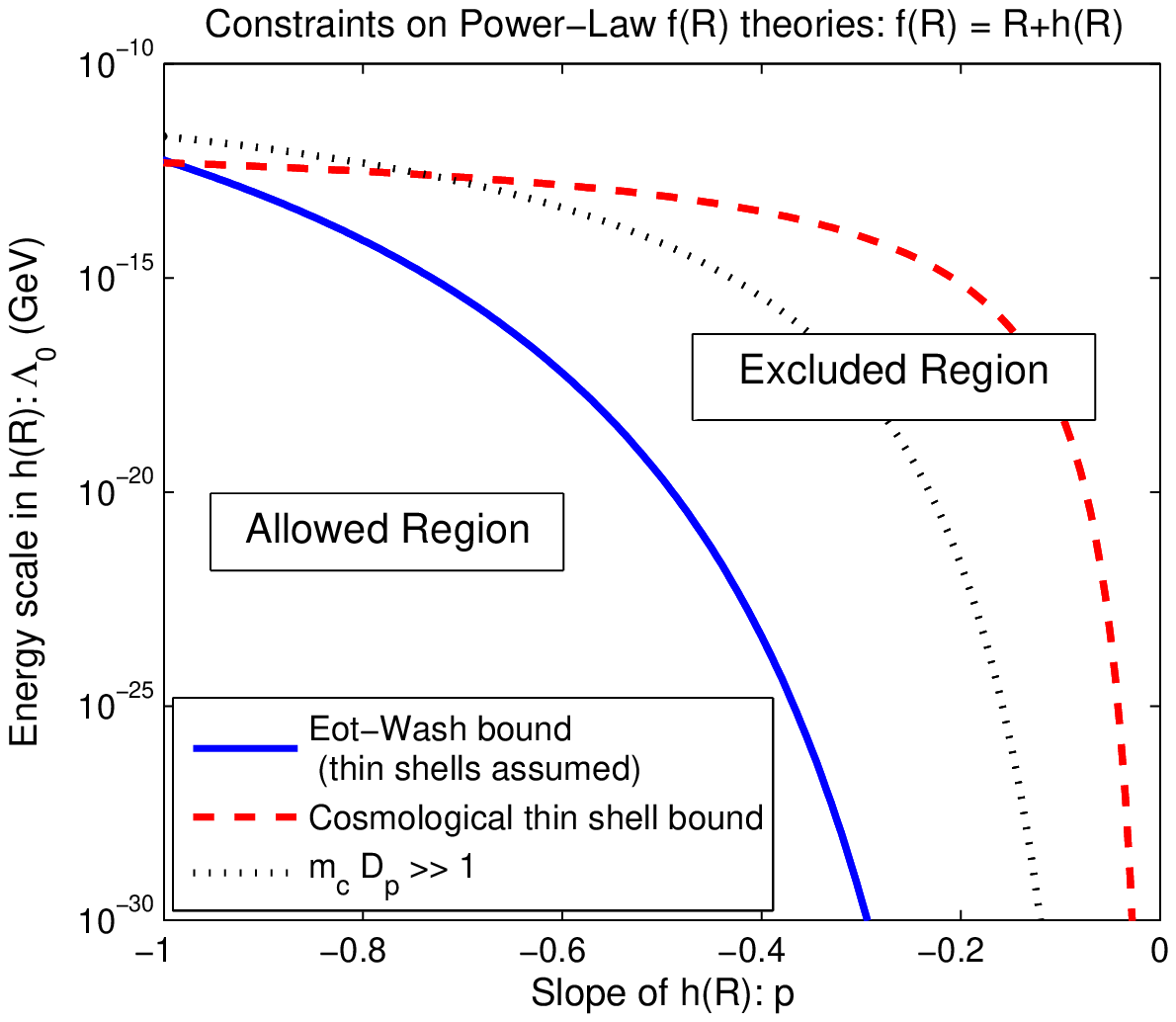}
\includegraphics*[width=6.5cm]{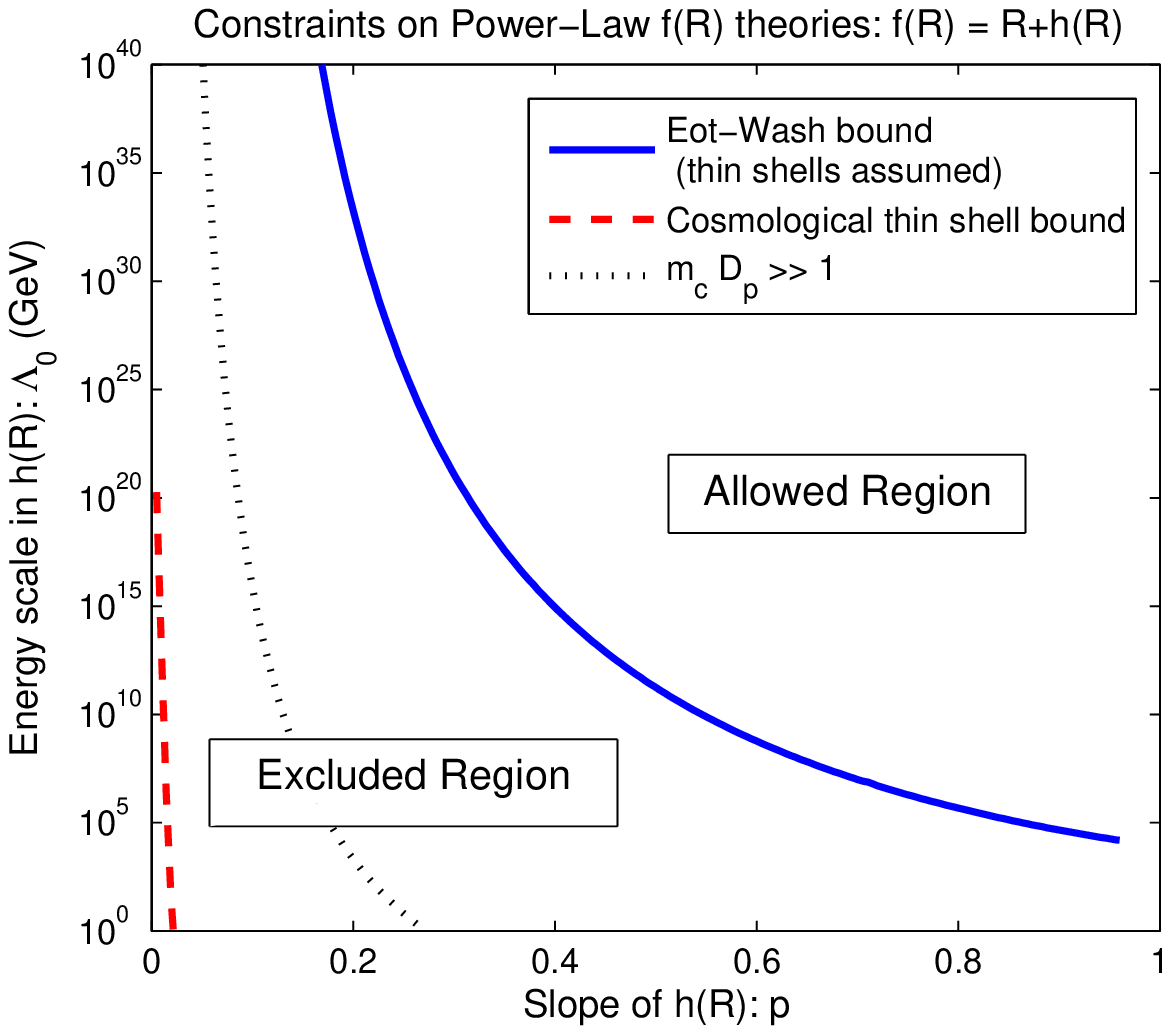}
\caption[]{E\"{o}t-Wash constraints (thick solid blue line) on
$f(R)$ gravity theories with $f(R) = R + h(R)$ where $h(R) =
\bar{R} (R/\bar{R})^{p+1}$; $\bar{R}=\Lambda_0^4/\mpl^2$ and $-5 <
p < -1$, $-1 < p < 0$ and $0 < p < 1$. For this constraint we have
assumed that the test bodies have thin-shells (which is necessary
to avoid local tests). We have also shown : (1) the
cosmological thin shell constraint (thick red dashed line) for
test bodies in the laboratory derived in Section \ref{sec:cosmo},
(2) the naive constraint (thick black dotted line) one could
derive by simply requiring that, inside the test bodies, the mass
of the chameleon at the minimum of its effective potential,
$m_c$is large compared with the length scale of the body, $D_{p}$.
This was the constraint considered in Ref. \cite{Staro}. For all
such theories we see that the correctly evaluated constraint
provided by the E\"{o}t-Wash experiment \cite{EotWash} is stronger
than both this na\"{i}ve constraint and the cosmological
thin-shell bound for all for $p \gtrsim -1$. The $m_c D_{\rm p}
\gg 1$ constraint never provides the strongest constraint. }
\label{fig1}
\end{figure}

Thus we define $x(d) = m_b d/a_p$ and $x_{\rm max} = {\rm min}(m_b r_{\rm h}/a_p, 1)$ and find:
\begin{eqnarray}
I(d, r_{\rm h}) \approx && \frac{p^2 \mpl^2 K_{p} \bar{R}^{1/2}}{2} \left(\frac{m_b}{\bar{R}^{1/2}a_p}\right)^{\frac{1+3p}{1-p}}  \\ &&\times \left \lbrace\left[H_{\rm p}(x(d)) - H_{\rm p}(x_{\rm max})\right] \right. \nonumber \\ &&\left.+ \frac{R_{c}^2 (x_{\rm max}-x)}{4 R_b^2 C(m_b r_{\rm p})}\mathcal{E}_{0}\right\rbrace, \nonumber
\end{eqnarray}
where
\begin{eqnarray}
H_{\rm p}(x) = &&\frac{1}{(1+p)}\left[\frac{1-p}{1+3p} x^{-\frac{1+3p}{1-p}} - \frac{x}{p} \right. \nonumber\\ &&\left. + \frac{1-p^2}{p(1-3p)} x^{\frac{1-3p}{1-p}}\right] .\nonumber
\end{eqnarray}
With this formula we are able to evaluate the E\"{o}t-Wash
constraint for all theories with $h(R) \propto R^{p+1}$.  We do
this further below. However, we discuss first the cosmological
thin-shell constraint on these theories.
\subsubsection{Cosmological Constraints}
On cosmological scales, the field $\phi$ is stuck at the minimum
of the effective potential provided $m_{\phi}^2 / H^2 \gg 1$ which
becomes:
$$
\frac{4\Omega_{\rm de} + \Omega_{\rm m}}{R h^{\prime \prime}(R)}  \gg 1.
$$
If $h(R) \propto R^{p+1}$ this becomes:
\begin{equation}
\frac{4\Omega_{\rm de} + \Omega_{\rm m}}{\vert p h^{\prime}(R) \vert} \gg 1. \label{masscon}
\end{equation}
The cosmological thin-shell constraint requires that:
$$
\frac{\beta \vert \Delta \phi \vert}{\mpl} \lesssim 10^{-4},
$$
where $\Delta \phi$ is the difference between the value of $\phi$
cosmological and the value of $\phi$ at the minimum of the
effective potential in a region with density $\mathcal{O}(1)\,{\rm
g}\,{\rm cm}^{-3}$.   This generally implies that cosmologically
$\beta \vert \phi \vert / \mpl \lesssim 10^{-4}$ and
$\frac{1}{2}\vert h^{\prime}(R) \vert \lesssim 10^{-4}$.  It is
clear then that Eq. (\ref{masscon}) holds provided $p \times
10^{-4} \ll 1$ and so for $\mathcal{O}(1)$ values of $p$, we are
always in the region where $m_{\phi}^2/H^2 \gg 1$ and $\phi$
lies close to the minimum of its effective potential.

To leading order we take $V_{,\phi} \approx -\beta \rho_{\rm matter}/\mpl$
and, defining $p = \ln a$, where $a$ is the FRW scale factor in the Jordan frame and $\Phi  = e^{-2\beta \phi/\mpl}$, we find that:
\begin{equation}
\frac{\Phi_{p}}{\Phi} \approx -\frac{3\Omega_{\rm m} H^2}{m_{\phi}^2} = -3f_0 Rh^{\prime \prime}(R) \ll 1,
\end{equation}
where $f_0 = \Omega_{\rm m}/(\Omega_{\rm m} + 4\Omega_{\rm de})$
\begin{equation}
\frac{\Phi_{pp}}{\Phi} \approx 9f_0 R h^{\prime \prime}\left[1 + \frac{f_0 R h^{\prime \prime \prime}(R)}{h^{\prime \prime}(R)}\right].
\end{equation}
Today from Eq. (\ref{weffeqn}), with $\Omega_{m}^{\rm eff} = \Omega_m$, $\Omega_{\rm de}^{\rm eff} = \Omega_{\rm de} \approx 1- \Omega_m$ we have $f_0 \approx \Omega_{\rm m}/(4-3\Omega_{\rm m})$:
\begin{equation}
(1+w_{\rm de}^{\rm eff})\Omega_{\rm de} \approx 3f_0 R h^{\prime \prime}(R)\left[\frac{4}{3} + \frac{f_0 R h^{\prime \prime \prime}(R)}{h^{\prime \prime}(R)} + \frac{\Omega_{\rm m}}{2}\right],
\end{equation}
and so for $\Omega_{\rm m} = 0.24$ and $h(R) \propto R^{p+1}$ we have:
$$
\vert 1+w_{\rm de}^{\rm eff} \vert \Omega_{\rm de}^{\rm eff} \approx 0.32\vert p h^{\prime}(R)\vert \left\vert 1 + 0.050(p-1)\right \vert.
$$
The cosmological thin-shell constraint ensures  that cosmologically $\vert h^{\prime}(R) \vert \lesssim 10^{-4}$ today and so:
$$
\vert 1+w_{\rm de}^{\rm eff} \vert \Omega_{\rm de}^{\rm eff}  \lesssim 3.2 \vert p \vert \left\vert 1 + 0.050(p-1)\right \vert \times 10^{-5}.
$$

\subsubsection{Collected Constraints}
We will consider now how the E\"{o}t-Wash data, when
thin-shells are assumed, constrains the properties of power-law
$f(R)$ theories. It should be stressed that in the absence of
thin-shells, the E\"{o}t-Wash would automatically rule out these
theories.

Defining $\bar{R} = \Lambda_{0}^4/\mpl^2$ we have plotted the
E\"{o}t-Wash constraints on $\Lambda_{0}$ for $-5 < p < 1$ in Fig.
\ref{fig1} as a thick (blue) solid line. The cosmological
thin-shell constraint is shown as a thick (red) dashed line. For
theories with $0 < p < 1$ we find a lower bound on $\Lambda_{0}$
and for theories with $p < 0$ we recover an upper bound.   We also
show, as a thick (black) dotted line, the na\"{i}ve constraint on
the parameters that one would find by simply requiring that the
chameleon mass at the minimum of the effective inside the plate,
$m_{c}$, is large compared to the plate thickness $D_{p} =
0.997\,{\rm mm}$.  It is a commonplace assumption in the
literature (see e.g. \cite{Staro, kaloper}) that assuming $m_{c} D_{\rm p} \gg
1$ (where more generally $D_{\rm p}$ would be the length scale of
the test body) is enough to satisfy local tests of gravity. It is
clear from the plots that this na\"{i}ve bound never provides the
tightest constraint on the parameters of the theory.
\begin{figure}[htb!]
\includegraphics*[width=6.5cm]{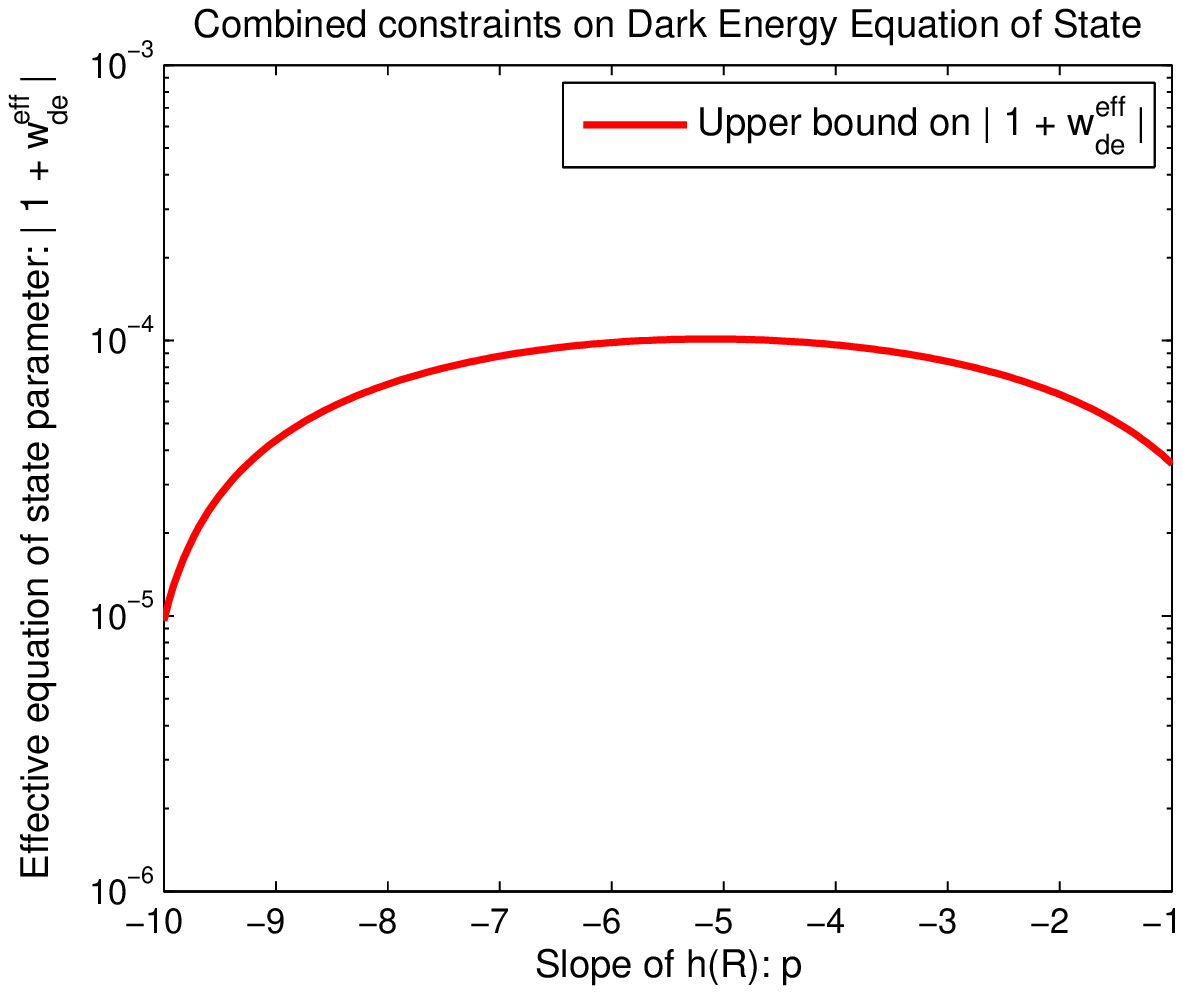}
\includegraphics*[width=6.5cm]{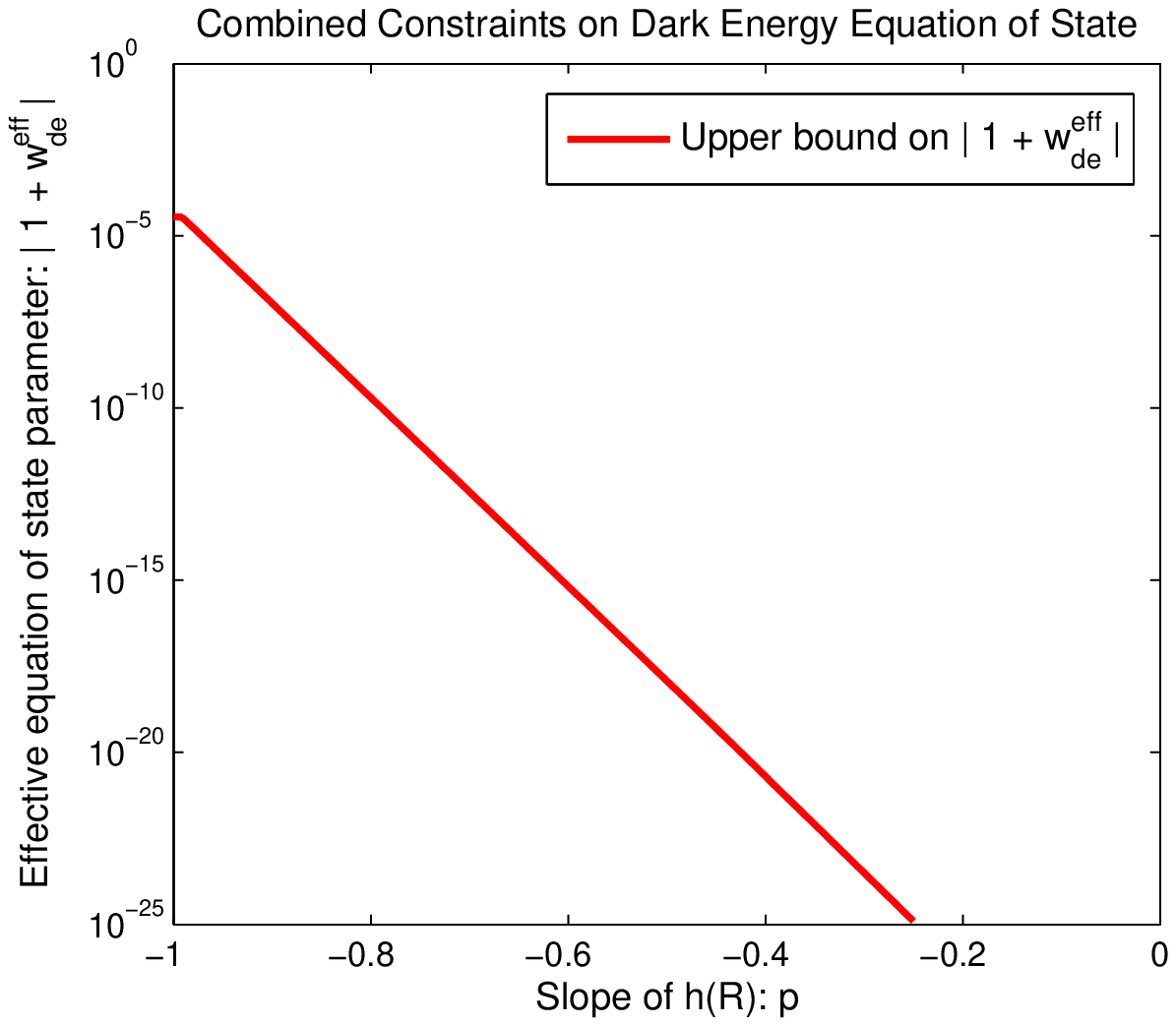}
\includegraphics*[width=6.5cm]{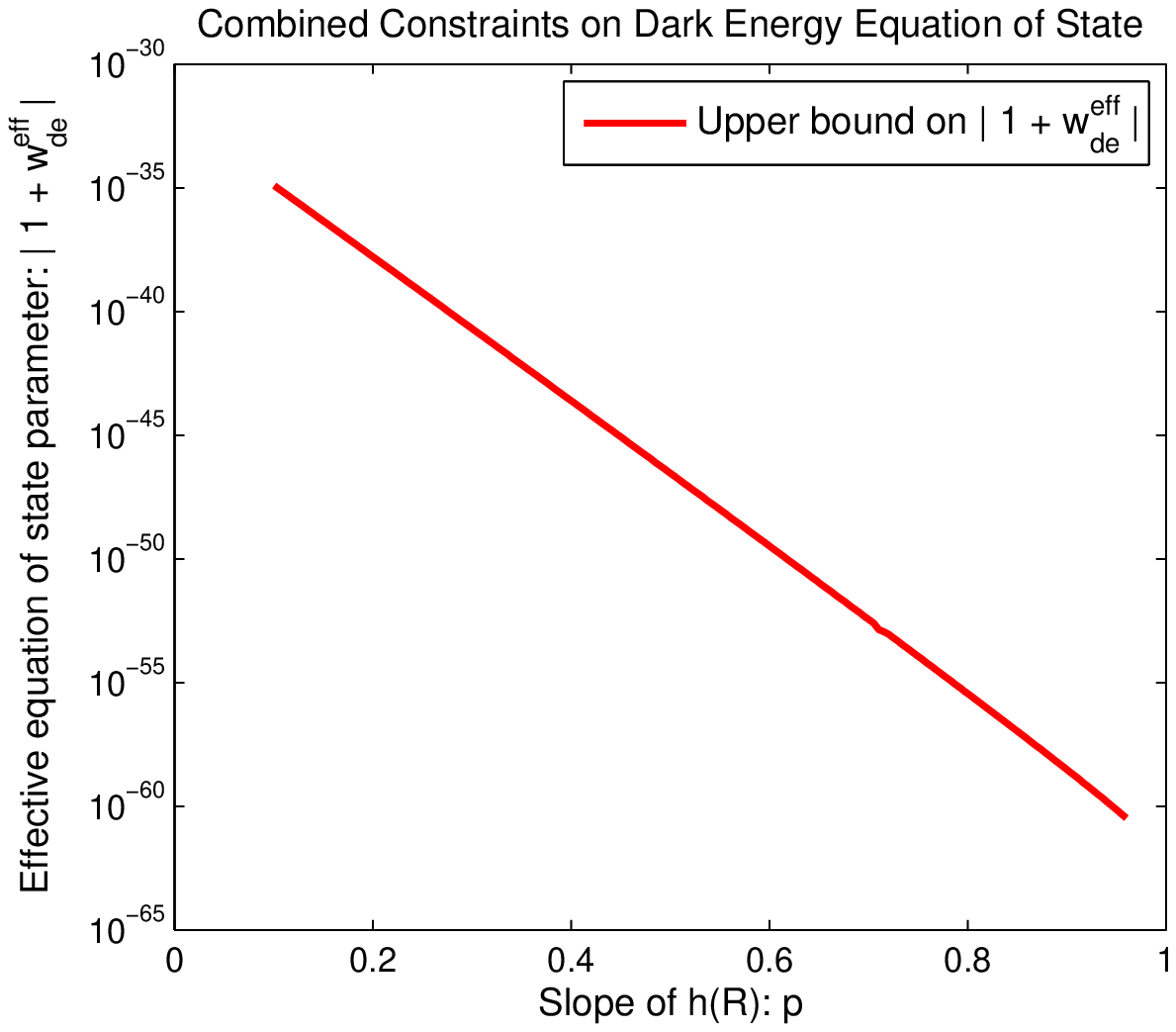}
\caption[]{Combined E\"{o}t-Wash constraints on the effective dark energy equation of state parameter produced by $f(R)$ gravity theories with $f(R) = R + h(R)$ where $h(R) = \bar{R} (R/\bar{R})^{p+1}$.   $\bar{R}=\Lambda_0^4/\mpl^2$ and $-5 < p < -1$, $-1 < p < 0$ and $0 < p < 1$.  These constraints have been derived by requiring both that the E\"{o}t-Wash test masses have thin-shells and by requiring that the chameleonic torque produced between the two thin-shelled test masses is small enough to have avoided detection to date. We see that in all cases we have $\vert 1 + w_{\rm de}^{\rm eff}\vert < 10^{-4}$ today. As a result, the late time cosmology of any viable theory would be virtually indistinguishable from that described by the $\Lambda$CDM model}
\label{fig2}
\end{figure}

The constraints on $\bar{R}$ constrain the equation of state parameter of the dark energy described by the $f(R)$ theory.  Taking $\Omega_m = 0.23$ today, we plot the collected constraints on the effective Jordan frame equation of state parameter (as defined in Section \ref{sec:cosmo}) in Fig. \ref{fig2}.  We see that at the current epoch $\vert 1 + w_{\rm de} \vert < 10^{-4}$ for all $\mathcal{O}(1)$ values of $p$ with the largest values occurring for $p < -1$, and hence the late time cosmology produced by any viable theory would be observationally indistinguishable from that described by the standard $\Lambda$CDM model.

\section{Conclusions}
In recent years, modifications of General Relativity have been suggested as a
possible explanation for the observed accelerated expansion of the universe. A
popular class of models are the so--called $f(R)$ theories.
While cosmologically viable theories can be found, local constraints on such
theories have to be worked out, since the gravitational sector is modified,
which could result in unacceptable deviations from Newton's law of gravity.

In this paper we have constrained $f(R)$ theories, using the well
known equivalence between these and scalar-tensor theories. For an
$f(R)$ theory to be consistent with both cosmology and local
gravity experiments, the equivalent scalar-tensor theory must be a
chameleon field theory. We have shown that the requirement of the
thin-shell mechanism at work in E\"ot-Wash experiments results in
an equation of state for dark energy very near to that of a
cosmological constant. Thus, viable $f(R)$ models (those which are
compatible with local experiments) behave on cosmological scales
similarly to the standard $\Lambda$CDM model and deviations are
expected only on very small (sub-galactic) scales. The expected
deviations from the cosmological constant equation of state $w=-1$ now
in viable $f(R)$ theories are unmeasurably small (at least with
current technologies). As examples, we have studied $f(R)$
theories with logarithmic potentials (based on \cite{kaloper}
for a fixed coupling $\beta=1/\sqrt 6$) as well as power-law
potentials (such as those presented in \cite{Hu2,Staro}). The
former are ruled out by local gravitational tests, while there is
still room for the latter models.

To conclude, while on cosmological scales viable $f(R)$ theories
behave like $\Lambda$CDM, deviations are expected on scales
which could be large enough to be within the reach of next
generation galaxy surveys \cite{spergel}. Hopefully, future
measurements of the dark matter distribution on those scales can be used to find
such deviations from the standard $\Lambda$CDM model. For this,
a detailed understanding of galaxy formation is necessary,
including an understanding of both the dynamics of baryons as
well as that of dark matter in $\Lambda$CDM and $f(R)$/chameleon
theories.

\vspace{0.5cm}

\noindent{\bf Acknowledgements:} CvdB, ACD and DJS are supported by STFC. We are grateful
to Lisa Hall for a careful reading of the manuscript and to Nemanja Kaloper for useful
comments.  DJS would like to thank  Kotub Uddin for helpful discussions, and D. J. Kapner for discussions and providing further details about the E\"{o}t-Wash experiment.

\appendix

\section{The force between two plates}
\label{appA}
In previous works \cite{chamstrong,chamcas}, the chameleonic force per
unit area between two parallel plates was calculated and found to be:
$$
\frac{F_{\phi}}{A} = V^{\prime}(\phi_{0}) - V^{\prime}(\phi_{b}) + \frac{\beta \rho_{b}}{\mpl}(\phi_0 - \phi_b),
$$
where $\phi_0$ depends on $d$. We assume that both plates have
radius $R$ and thickness $D$ and that $D \ll R$.  However as these
calculations  treat the plates as being infinite, they required
for consistency that:
\begin{itemize}
\item \emph{either}, for an isolated plate, $\phi \rightarrow
\phi_b$ at a distance, $d \ll R$, from the plate so that the
infinite plate approximation was still valid, \item \emph{or}, the
precise value of $\phi_b$ was not important when the plates were
separated by a distance $d \ll R$.  This means that provided
$V^{\prime}(\phi_0)/V^{\prime}(\phi^{\ast}) \ll 1$, one could
replace $\phi_b$ in the above expression by $\phi^{\ast}$ without
altering the prediction for $F_{\phi}/A$ greatly. In these cases,
the behaviour of $\phi$ far from the plates, where the infinite
plate approximation is invalid, would be unimportant.
\end{itemize}
These approximations held for all of the chameleon theories
considered in Refs. \cite{chamstrong,chamcas}, however in this
work we consider a wider range of theories, and it is often the
case that both  of these assumptions fail to hold.  In this
appendix, we therefore derive an improved version of the force
formula.

Outside of a body in a region where the background density is $\rho_b$, the chameleon field obeys:
$$
\nabla^2 \phi = V_{,\phi}(\phi) + \frac{\beta \rho_b}{\mpl}.
$$
Consider this equation near one of the circular surfaces of a cylindrical plate, of uniform density, and with radius $R$ and thickness $D \ll R$.  Defining $z$ to be the distance from one of the circular surfaces of the plate we have:
\begin{equation}
\frac{\partial^2 \phi}{\partial z^2} = V_{,\phi}(\phi) - V_{,\phi}(\phi_b) - \frac{1}{r}\frac{\partial }{\partial r} \left(r \frac{\partial \phi}{\partial r}\right). \label{MainEqn}
\end{equation}

We begin by consider the case where only one plate is present. Here $\phi\rightarrow \phi_b$, $\partial \phi / \partial z \rightarrow 0$ as $z \rightarrow \infty$. Integrating Eq. (\ref{MainEqn}) with these boundary conditions give:
\begin{eqnarray}
\frac{1}{2} \left(\frac{\partial \phi}{\partial z}\right)^2 = &&V(\phi) - V(\phi_b) - V_{,\phi}(\phi_b)(\phi-\phi_b) \nonumber \\
&& + \int_{z}^{\infty}\frac{1}{r}\frac{\partial }{\partial r} \left(r \frac{\partial \phi}{\partial r}\right) \frac{\partial \phi}{\partial z} \dd z. \label{orgEq}
\end{eqnarray}
We solve this approximately by assuming that for $z < z^{\ast}$
the $z$-dependence of the $r$-derivative terms is weak compared to
that of the potential terms, and that for $z
> z^{\ast}$, the non-linear terms in the potential, i.e. terms
that depend on 3rd or higher derivatives of $V$, are sub-dominant.
In  $z > z^{\ast}$,  we have $\phi \approx \tilde{\phi}$ where:
 \begin{eqnarray}
 \frac{1}{2} \left(\frac{\partial \tilde{\phi}}{\partial z}\right)^2 = && \frac{1}{2}m_{b}^2(\tilde{\phi}-\phi_b)^2 , \label{plEqn1} \\
&& + \int_{z}^{\infty}\frac{1}{r}\frac{\partial }{\partial r} \left(r \frac{\partial \tilde{\phi}}{\partial r}\right) \frac{\partial \tilde{\phi}}{\partial z} \dd z. \nonumber
 \end{eqnarray}
 or equivalently:
\begin{equation}
\frac{\partial^2 \tilde{\phi}}{\partial z^2} \approx m_{b}^2(\tilde{\phi} - \phi_b)  - \frac{1}{r}\frac{\partial }{\partial r} \left(r \frac{\partial \tilde{\phi}}{\partial r}\right). \nonumber
\end{equation}
Assuming that the plate is thin ($D \ll R$), and solving Eq.
(\ref{plEqn1}), we find that for the $z > z^{\ast}$ and along
$r=0$:
$$
\tilde{\phi} - \phi_b \propto \left(e^{-m_{b} z}-e^{-m_{b}r}\right),
$$
and here $z$ is the distance from the plate surface. It follows
that for $z \ll R$ we have:
\begin{equation}
\int_{z}^{\infty}\frac{1}{r}\frac{\partial }{\partial r} \left(r \frac{\partial \tilde{\phi}}{\partial r}\right) \frac{\partial \tilde{\phi}}{\partial z} \dd z  = m_{b}^2 (\tilde{\phi}-\phi_b)^2\frac{(e^{m_b R} - 1/2)}{(e^{m_b R} -1)^2}. \label{rgrad}
\end{equation}
In $z < z^{\ast}$, we have assumed that the $z$-dependence of the $r$-gradient terms is relativity weak. We therefore approximate the $r$ gradient terms in Eq. (\ref{orgEq}) using the $z > z^{\ast}$ solution i.e. we approximate them using Eq. (\ref{rgrad}) with $\tilde{\phi} \rightarrow \phi$.  For $z \ll R$ we then have
\begin{eqnarray}
\frac{1}{2}\left(\frac{\partial \phi}{\partial z}\right)^2 &&\approx V(\phi) - V(\phi_b) + V_{,\phi}(\phi_b)(\phi-\phi_b) \nonumber \\
&& + m_{b}^2 (\phi-\phi_b)^2\frac{(e^{m_b R} - 1/2)}{(e^{m_b R} -1)^2}.
\end{eqnarray}
The above equation also holds approximately in the $z> z^{\ast}$ region, provided $z \ll R$, and so provides an approximation to the evolution of $\phi$ everywhere when $z \ll R$.   In particular we see that on the surface of the plate, at $z=0$, where say $\phi = \bar{\phi}_{s}$:
\begin{eqnarray}
\frac{1}{2}\left(\frac{\partial \phi}{\partial z}\right)^2  &&\approx V(\bar{\phi}_s) - V(\phi_b) + V_{,\phi}(\phi_b)(\bar{\phi}_s-\phi_b)\nonumber\\
&& + m_{b}^2 (\bar{\phi}_{s}-\phi_b)^2\frac{(e^{m_b R} - 1/2)}{(e^{m_b R} -1)^2}.\label{outeqn1}
\end{eqnarray}

We assume that the plate has a thin-shell, so that deep inside it $\phi \rightarrow \phi_c$ where:
$$
V_{,\phi}(\phi_c) = -\frac{\beta \rho_c}{\mpl}.
$$
Provided the shell is thin, we can treat the system as being essentially 1 dimensional \cite{chamstrong,chamcas} and so:
\begin{equation}
\frac{1}{2}\left(\frac{\partial \phi}{\partial z}\right)^2 =  V(\phi) - V(\phi_c) - V_{,\phi}(\phi_c)(\phi-\phi_c). \label{inEqn}
\end{equation}
Thus by evaluating and equating the left hand sides of Eqs. (\ref{outeqn1}) and (\ref{inEqn}) at the surface we find:
\begin{eqnarray}
&& V(\phi_c) - V(\phi_b) - V_{,\phi}(\phi_c)(\phi_c - \phi_b) + (V_{,\phi}(\phi_c) \nonumber \\ &&- V_{,\phi}(\phi_b))(\bar{\phi}_s -\phi_b) \nonumber \\ &&+ m_b^2 C(m_b R) (\bar{\phi}_s-\phi_b)^2 = 0,\nonumber
\end{eqnarray}
where
$$
C(m_b R) = \frac{(e^{m_b R} - 1/2)}{(e^{m_b R} -1)^2}.
$$
Thus:
\begin{eqnarray}
\bar{\phi}_{s} -\phi_b &&= \frac{(V_{,\phi}(\phi_b) - V_{,\phi}(\phi_c))}{2 m_{b}^2 C(m_b R)}\left[1\right. \nonumber  \\ &&\left. - \sqrt{1 + 4C(m_b R)D(\phi_c, \phi_b)}\right] \label{barphis}
\end{eqnarray}
where
$$
D(\phi_c, \phi_b, m_{b}R) = \frac{m_{b}^2\left[V(\phi_b) - V(\phi_c) - V_{,\phi}(\phi_c)(\phi_b - \phi_c)\right]}{(V_{,\phi}(\phi_b) - V_{,\phi}(\phi_c))^2}.
$$

We now consider the force between two parallel plates.  This derivations make uses of results found in Refs. \cite{chamstrong} and \cite{chamcas}, and proceeds along roughly similar lines.

In between two parallel plates with radius $R$ and with separation $d \ll R$ in the $z$-direction, the chameleon field obeys:
\begin{equation}
\frac{\partial^2 \phi}{\partial z^2} = V_{,\phi}(\phi) - V_{,\phi}(\phi_b) \label{phiEqnapp}
\end{equation}
For simplicity we treat the plates as having the same composition. This assumption was dropped in Ref. \cite{chamstrong}, however, it was also shown there that for most purposes the assumption provides an excellent approximation. This is because the chameleonic force generally exhibits very little composition dependence \cite{chamstrong}.  We define $z =0$ to be the surface of one of the plates, and $z=d$ to be the facing surface of the second plate.   The system is symmetric and so $\dd \phi / \dd z = 0$ at $z=d/2$.  We define $\phi(z=d/2) = \phi_0(d)$.  A formulae for $\phi_{0}(d)$ have been provided in Ref. \cite{chamcas}.  Integrating Eq. (\ref{phiEqnapp}) we have:
\begin{equation}
\frac{1}{2}\left(\frac{\partial \phi}{\partial z}\right)^2 =  V(\phi) - V(\phi_0) - V_{,\phi}(\phi_b)(\phi-\phi_0). \label{outEqn1}
\end{equation}
Following Ref. \cite{chamcas}, inside either plate, Eq. (\ref{inEqn}) holds. By equating both Eqs. (\ref{outEqn1}) and (\ref{inEqn}) at the surface of one of the plates, where $\phi = \phi_{s}$ say, we find:
$$
\phi_{s} = \frac{\left[V(\phi_c) - V(\phi_0) + V_{,\phi}(\phi_b)\phi_0 - V_{,\phi}(\phi_c)\phi_c\right]}{V_{,\phi}(\phi_b) - V_{,\phi}(\phi_c)}.
$$

In Ref. \cite{chamcas} it was shown that the attractive chameleonic force unit area between two thin-shelled plates is given by $-V_{,\phi}(\phi_c) (\phi_s-\bar{\phi}_{s})$, and if, as is usually the case, the plates are much denser than their environment so that $V_{,\phi}(\phi_c)/V_{,\phi}(\phi_b) = \rho_c/\rho_b \gg 1$, we have:
\begin{equation}
\frac{F_{\phi}}{A} = V(\phi_0) - V(\phi_{b}) - V_{b}^{\prime}(\phi_0 - \phi_{b})  + m_{b}^2 C(m_{b}R)(\bar{\phi}_{s} -\phi_b)^2. \label{fEqnApp}
\end{equation}
This coincides with the formulae found in Refs. \cite{chamstrong,chamcas} when $m_{b}R \gg 1 \Rightarrow  C(m_b R) \approx 0$, or more generally whenever the final term is small compared with the other terms, which is when $C(m_b R)D(\phi_c,\phi_b) \ll 1$. $\bar{\phi}_s - \phi_b$ is given by Eq. (\ref{barphis}).

When $m_b R \ll 1$, we have $C(m_b R) \approx 1/2m_b^2 R^2 \gg 1$. If this is the case we also have $C(m_b R)D(\phi_c, \phi_b) \gg 1$:
$$
m_{b}^2 C(m_b R)(\bar{\phi}_s -\phi_b)^2 \approx  V(\phi_b) - V(\phi_c) - V_{,\phi}(\phi_c)(\phi_b -\phi_c),
$$
and then in this limit Eq. (\ref{fEqnApp}) becomes:
$$
\frac{F_{\phi}}{A} = V(\phi_0) - V(\phi_c) - V_{c}^{\prime}(\phi_{b}-\phi_c) - V_{b}^{\prime}(\phi_0 - \phi_{b}).
$$

\section{Chameleon Mass Between Two Plates}
\label{appB}
In this appendix we generalize the calculation of the chameleon mass between two parallel plates, as performed in Refs. \cite{chamstrong,chamcas}, to include the wider range of chameleon theories considered here.

In between two parallel plates with, say, a circular cross section, in the $x-y$ plane, with radius $r_{\rm p}$, and a separation, $d$, in the $z$-direction where  $d \ll r_{\rm p}$, Eq. (\ref{eqnPhi}) for $\phi$ simplifies to be essentially one dimensional $\square \rightarrow \frac{\dd^2 }{\dd z^2}$:
\begin{equation}
\frac{\dd^2 \phi}{\dd z^2} = V_{,\phi}(\phi) - V_{,\phi}(\phi_b),
\end{equation}
where $\phi_b$ is the background value of $\phi$.  We define $\phi_0$ to be the value of $\phi$ when $\dd \phi /\dd z =0$, which will occur midway between the two plates when $z = d/2$.

Thus integrating the $\phi$ equation we find:
\begin{equation}
\frac{1}{2}\left(\frac{\dd \phi}{\dd z}\right)^2 = V(\phi) -V(\phi_0) - V_{,\phi}(\phi_b)(\phi - \phi_b). \nonumber
\end{equation}
In this work we have considered power law potentials where $V \propto \epsilon p (-\epsilon\phi/(p+1))^{\frac{p+1}{p}}$ and where $\epsilon = {\rm sgn}(p(p+1))$ and $p < 1$.  For these potentials we have $m_{\phi}^2 = V_{,\phi \phi} = (p+1)/p^2 V(\phi)/\phi^2$. Thus, defining $Y = \phi/\phi_0$ and $m_{0} = m_{\phi}(\phi_0)$ we have:
\begin{eqnarray}
&&\frac{1}{2} \left(\frac{\dd Y}{\dd z}\right)^2 = \frac{p^2 m_{0}^2}{(p+1)} \left[Y^{\frac{p+1}{p}}\right. \label{appEqn4} \\ &&\left.  - 1 - \frac{p+1}{p}\left(\frac{\phi_b}{\phi_0}\right)^{\frac{1}{p}} (Y - 1) \frac{V(\phi_0)}{\phi_0}\right].  \nonumber
\end{eqnarray}
Now
$$
R \propto \phi^{1/p},
$$
and so defining $R_0 = R(\phi_0)$ we have $(\phi_b/\phi_0)^{1/p} = R_b /R_0$. Thus when $R_b / R_0 \ll 1$, the last term in Eq. (\ref{appEqn4}) is very small and can be dropped.  Working in this limit we have:
$$
\frac{1}{2} \left(\frac{\dd Y}{\dd z}\right)^2 \approx \frac{p^2 m_{0}^2}{(p+1)} \left[Y^{\frac{p+1}{p}} - 1\right].
$$
Now on the surface of the plates $\phi \sim \mathcal{O}(\phi_c)$ where $\phi_c$ is the value of $\phi$ inside the body, and assuming $R_c \gg R_0$ i.e. $m_{c} \gg m_{0}$, we can we treat $(\phi/\phi_{0})^{p}$ as becoming very large as $z \rightarrow 0$ (i.e. as we approach the surface of the plate). We define $X = Y^{-p} = (\phi/\phi_0)^{-p}$ and then in the limit $m_{b} \ll m_0 \ll m_c$ we have:
$$
\left(\frac{\dd X}{\dd z}\right)^2 \approx \frac{2m_{0}^2 X^{2+2/p}}{(p+1)} \left[X^{-\frac{p+1}{p^2}} - 1\right].
$$
Integrating this equation and using $X=0$ at $z=0$, $X =1$ at $z=d/2$ we have for $p \leq -1$:
\begin{equation}
\frac{m_{0} d}{\sqrt{2\vert p+1 \vert }} = \frac{p^2}{\vert p+1\vert} B\left(\frac{1}{2}, \frac{p)}{(1+p)}\right), \nonumber
\end{equation}
which simplifies to:
\begin{equation}
m_{0} d = \sqrt{\frac{2}{\vert 1+p\vert}} p^2 B\left(\frac{1}{2}, \frac{p}{(1+p)}\right).
\end{equation}
If $-1 \leq  p \leq 1$ then we find:
\begin{equation}
m_{0}d = \sqrt{\frac{2}{\vert 1+p\vert}} B\left(\frac{1}{2}, \frac{1-p}{2(1+p)}\right).
\end{equation}

\end{document}